\documentclass[extra,onecolumn,fleqn,mreferee]{gji}
\usepackage[fleqn]{amsmath}
\usepackage{amssymb}
\usepackage{bm}
\usepackage{ulem}
\usepackage{graphicx}
\usepackage{subcaption}
\usepackage{xcolor}
\usepackage{microtype}
\usepackage{mathtools} 
\usepackage{upgreek}
\usepackage{hyperref}
\usepackage{lipsum}
\usepackage{xifthen} 

\usepackage{general_sty}
\usepackage{mathematical_shorthands}
\usepackage{continuum_mechanics_definitions}
\usepackage{rotational_sty_2}

\graphicspath{{./figures}}

\numberwithin{equation}{section}

\begin{document}

\title{On the elastodynamics of rotating planets}

\author[M.A. Maitra \& D. Al-Attar]
   {Matthew Maitra$ ^{1} $ and David Al-Attar$ ^{2} $ \\
  $ ^1 $Institut f\"{u}r Geophysik,
    ETH Z\"{u}rich,\\
    Sonneggstrasse 5,
	Z\"{urich} 8092, 
	Switzerland\\
  $ ^2 $Bullard Laboratories,
    Department of Earth Sciences,
    University of Cambridge,\\
	Madingley Rise, 
	Madingley Road, 
	Cambridge CB3 0EZ, UK
  }

\date{\today}

\maketitle

\begin{summary}
Equations of motion are derived for (visco)elastic, self-gravitating, and variably-rotating planets. The equations are written using a decomposition of the elastic motion that separates the body's elastic deformation from its net translational and rotational motion as far as possible. This separation is achieved by introducing degrees of freedom that represent the body's rigid motions; it is made precise by imposing constraints that are physically motivated and should be practically useful. In essence, a Tisserand frame is introduced exactly into the equations of solid mechanics. The necessary concepts are first introduced in the context of a solid body, motivated by symmetries and conservation laws, and the corresponding equations of motion are derived. Next, it is shown how those ideas and equations of motion can readily be extended to describe a layered fluid--solid body. A possibly new conservation law concerning inviscid fluids is then stated. Thereafter the equilibria and linearisation of the fluid--solid equations of motion are discussed, along with new equations for use within normal-mode coupling calculations and other Galerkin methods. Finally, the extension of these ideas to the description of multiple, interacting fluid--solid planets is qualitatively discussed.
\end{summary}


\section{Introduction}
\label{sec:intro}

The continuum mechanics of rotating, self-gravitating, viscoelastic, layered fluid--solid planets is relevant in studies of: seismology, co- and post-seismic deformation, Earth rotation, tides and glacial isostatic adjustment (GIA). Since the underlying equations can be formulated in different ways, it is natural to ask how best to do so.
The standard approach as summarised in Chapter 3 of
\citet[][`D\&T' hereafter]{dahlen1998theoretical} 
is to write the exact equations of finite elasticity
\citep[\eg][]{Coleman_1961,Coleman_1964,Biot_1965,Noll_1974,Truesdell2004}
with respect to a uniformly rotating reference frame whose origin is fixed in inertial space, to take a referential (`Lagrangian') view of particle motions and a spatial (`Eulerian') view of the gravitational potential, and then to linearise about equilibrium. 
This approach can be seen as a culmination of geophysical work that began in the late 1960s \citep{Backus_1967a,Dahlen_1972a} and was continued through the 1970s by \citet{Dahlen_1972b}, \citet{Smith_1974}, \citet{Dahlen_1975} and \citet{Woodhouse_1978}. 
In this paper we present a different approach: we formulate the equations of continuum mechanics using an entirely referential description and with respect to a dynamically evolving reference frame. This idea has been directly inspired by work on 
Euler-Poincar\'{e} reduction within geometric mechanics \citep[\eg][]{holm2009geometric,marsden2013introduction}.
Our approach clarifies the physical unity of Earth's rotation and elastodynamics, providing a general framework that bridges the respective theoretical approaches within  geodesy and seismology and that generalises both \citep[\cf][p.~104]{Smith_1977}.
We also emphasise the exact nonlinear formulation of the problem, in contrast to earlier studies that focused on linearised theories.

Before discussing the novel aspects of this work, it is worth emphasising that the final equations of motion we obtain take a broadly similar form to, and are not appreciably more complex than, those of conventional formulations (\eg D\&T, Chs.~2\&3). For example, we find a momentum equation almost identical to D\&T's, just with a dynamical angular velocity and associated Euler force, and subject to some constraints. This momentum equation is coupled to simple ODEs governing the evolution of the reference frame. There is therefore no real penalty incurred in using our reformulation, with near-identical numerical methods being applicable in both cases. On the other hand, as will be explained below, our equations have a greater range of applicability while offering certain conceptual advantages.

Our treatment of rotation is based upon an exact implementation of the \textit{Tisserand frame} \citep[\eg][]{munk1960rotation} within continuum mechanics. By definition, the deformation of a planet viewed from such a frame possesses no \textit{net} linear or angular momentum. Instead, the evolution of the frame 
captures any translational and rotational aspects of the planet's motion.
Although discussion and use of such a frame is not uncommon within observational studies, we are not aware of its exact implementation within the equations of solid mechanics (though see \citet{Rose_2017} for comparable ideas in the context of mantle convection).
In contrast to conventional formulations based on a steadily rotating reference-frame, the Tisserand frame simplifies the study of rotating bodies' elastodynamics. 
For example, consider how the melting of an ice-sheet will affect a planet's motion. Not only will the change in surface loading cause (visco)elastic deformation but, through changes in the planet's moment-of-inertia, there will also be an associated rotational perturbation. Viewed from a steadily rotating reference frame, such a rotational perturbation manifests as a secular growth in the displacement. From the Tisserand frame, however, that perturbation is absorbed by the frame's motion and there is no secular component to the planet's internal deformation. 
More generally, as discussed by \citet{Canavin_1977}, the Tisserand frame is preferable to a steadily rotating frame for bodies undergoing large rotational variations because the latter frame can make small internal deformations look larger than they are. 
Within conventional geophysical applications there is perhaps no great need to consider large rotational perturbations, but we reiterate that the use of the Tisserand frame does not entail major complications. Furthermore, such large perturbations are relevant within certain planetary science problems \citep[\eg][]{Cuk_2016,Lock_2018}, and there the use of a steadily rotating frame would be inconvenient within an exact theory and could lead to potentially significant errors upon linearisation.

As noted above, the standard geophysical approach to continuum mechanics mixes spatial and referential variables. 
The reason for this mixed formulation seems to be largely historical, perhaps motivated by a desire to obtain familiar-looking equations. Here, in contrast, we use a consistently referential description of the problem. While either approach is valid, we feel that a fully referential treatment is clearer even if initially less familiar. In the context of numerical work, a spatial formulation is often just as convenient as a referential one, but for certain methods \citep[\eg][]{Maitra_2019,Leng_2019} a referential formulation is essential.

A notable feature of this paper is its emphasis on the exact nonlinear equations of continuum mechanics.
We develop such equations in a form suitable for numerical implementation, taking the view that nonlinear rather than traditional linearised equations will be most useful in future numerical work. 
To give an example, recent work on modelling GIA has focused on the effect of lateral variations in Earth structure, including that of  realistic surface topography \citep[\eg][]{Latychev_2005}. The amplitude of such topography is, however, comparable to the surface deformation produced close to continental ice sheets, and so one could ask whether the use of linearised theory is appropriate. A similar point could be made in the context of co- and post-seismic deformation modelling \citep[\eg][]{Gharti_2019} if the cumulative effects of many earthquake cycles are considered. 
Importantly, approaching such problems using the full nonlinear equations should not entail unnecessary complications because with modern numerical methods the cost of solving a weakly nonlinear problem is similar to that of the corresponding linearised problem. Thus, if nonlinear effects do not matter then one might as well solve the nonlinear problem, while if such effects do matter then one is obliged to solve the nonlinear problem.

This paper places particular emphasis on symmetries and associated conservation laws. For this reason we start from \textit{Hamilton's principle of Least Action} and use \textit{Noether's theorem} to derive conservation laws. The Euclidean symmetries that underlie conservation of linear- and angular-momentum directly motivate our introduction of the Tisserand frame. We also discuss an infinite-dimensional symmetry-group associated with inviscid fluid regions, and use this to derive a novel conservation law. It is the latter symmetry that allows us to propose a new, fully referential approach to modelling quasi-static deformation in planets with a fluid core \citep[\cf][]{Dahlen_1974}. As a special case, we show that the standard potential formulation for the displacement within neutrally-stratified fluid regions \citep[\eg][]{Komatitsch_2002} can be derived from the conservation law. We anticipate that further study of this conservation law could help address complications related to core undertone modes \citep[\eg][]{Crossley_1975,Valette_1989b,Rogister_2009}.

An apparent drawback of using Hamilton's Principle is that it is only valid for non-dissipative materials. Nevertheless, Hamilton's Principle leads to equations of motion of precisely the same form as do derivations starting from, say, balance-laws or a weak formulation. The only practical difference between the two approaches is that the latter does not require the first Piola--Kirchhoff (FPK) stress to be derived from a strain-energy function. Instead, the existence of the FPK stress follows from \textit{Cauchy's theorem} \citep[e.g.][p.127]{marsden1994mathematical} and constitutive behaviour is imposed by relating the FPK stress to the deformation gradient history \citep[e.g.][]{Coleman_1961}. On a pragmatic level, all that this means is that we \textit{can} use this paper's equations of motion to model dissipative bodies, but we must just remember that the stress is not related to a strain-energy function. 

The paper is organised as follows. We begin in Section~\ref{sec:solid} by considering solid bodies. Then in Section~\ref{sec:fluid_solid} we show how Section~\ref{sec:solid}'s methods can readily be extended to fluid--solid bodies through the work of \citet[][`A18' henceforward]{Al_Attar_2018}. This presentation allows us to introduce the Tisserand frame in a simple context before considering complications due to fluid--solid boundaries. In Section~\ref{sec:multiple} we discuss qualitatively how the methods developed for a single fluid--solid body can be generalised to problems of multiple interacting bodies. We give two examples: first we discuss elasticity's exact effect on the Keplerian two-body problem; secondly, we suggest a new, exact approach to the Earth's quasi-rigid motions. Finally, in Section~\ref{sec:conc} we sum up the preceding work and speculate on paths forward.

\section{Hamilton's principle for a self-gravitating and rotating solid elastic planet}
\label{sec:solid}

Here we show how Hamilton's Principle is used to derive the equations of motion and study the symmetries of a solid elastic planet. In Section~\ref{sec:solid_foundations} we state this paper's initial version of Hamilton's principle, then discuss symmetries and conservation laws in Sections~\ref{sec:symmetries}~and~\ref{sec:solid_noether}. Section~\ref{sec:decomposition_solid} introduces this paper's main idea of separating elastic from rigid motion, then in Sections~\ref{sec:var_prin_solid_decomp} and \ref{sec:single_body_eoms} we restate the action accordingly and derive new equations of motion. Unless otherwise stated, \citet{marsden1994mathematical} is our source for standard results from continuum mechanics.

\subsection{Foundations}
\label{sec:solid_foundations}

\subsubsection{Kinematics}
\label{sec:elastic_body_kinematics}

To describe the deformation of an elastic body we require the concepts of a \textit{reference body} and a \textit{configuration}. 
A reference body is a compact subset \( \BB \subseteq \RThree \) with open interior and smooth boundary, \( \partial \BB\), whose 
points correspond bijectively to particles of the body. The form of the reference body encodes topological 
information (e.g. the body's connectedness) but is otherwise arbitrary. The possible states of the body are
described by mappings from \(\BB\) into \(\RThree\), with each such mapping known as a configuration. We require that
these configurations are smooth \textit{embeddings} of \(\BB\) into \(\RThree\), this meaning that such a mapping, \( \phib\), is smooth, 
injective,
and that there exists a smooth inverse mapping \(\phib^{-1}:\phib(\BB) \rightarrow \BB\) defined on its image. Physically, these conditions
mean that the body cannot be torn nor interpenetrate during its motion. We will write the set of all such embeddings as 
\(\EMB{}\), with this set forming the configuration space within our problem. Without dwelling on any technical details, 
we note that \(\EMB{}\) forms an infinite-dimensional manifold \citep[e.g.][]{Abraham_2012}; this geometric point of view is sometimes useful.   
In Section~\ref{sec:grav_variational} we will slightly modify this picture so as to facilitate a referential approach to self-gravity. Further changes are also necessary in Section~\ref{sec:fluid_solid_kinematics}  when we consider internal boundaries within the body across which the material parameters fail to be smooth. In both cases, the result is that the configuration space needs to be suitably generalised, but the following definitions remain valid.

As time progresses, the body adopts different configurations. Its \textit{motion} can then be viewed as
a time-parametrised curve within the configuration space, \(\EMB{}\), that we denote by
\begin{align}
\phib:\III \rightarrow \EMB{},
\end{align}
where \( \III = \braksq{0,T} \) is the time-interval of interest. We write \(\BB_{t}\) for the image of 
\(\BB\) under \(\phib(t) \). The configuration \(\phib(t)\) carries the particle labelled by \(\xx \in \BB\)
 to the point \(\yy = \phib(t)(\xx)\) in physical space. We can equivalently consider the motion as a mapping
 \begin{align}
   (\xx,t)   \mapsto \phib(t)(\xx),
 \end{align}
 which returns the position of the particle \(\xx\) at time \(t\). By a harmless abuse of notation, we
 will also denote this latter mapping, by \(\phib\), and hence write \(\phib(\xx,t) = \phib(t)(\xx)\). Here we see that the motion can be viewed as either 
 (i)  a function of time that takes values within a certain function space, or (ii)  a function of space and time that 
 takes values in \(\RThree\). Both perspectives are useful in different contexts, and the same holds for some other 
  quantities introduced below.

The \textit{velocity}, \(\vv\), is the time-derivative of the motion, with \(\vv(t)\) taking its values in the tangent space to \( \EMB{}\) based at \(\phib(t)\). Note that because
\( \RThree\) is flat, this latter space can be identified with the space of vector fields on \( \BB \). The velocity
of a particle, \(\xx\), at time, \(t\), can be written as \(\vv(t)(\xx)\) or equivalently, and more conventionally, as \(\vv(\xx,t)\).

For a configuration, \(\phib \in \EMB{}\), we can form its derivative, \(D \phib\), which is 
defined through the first-order Taylor expansion
\begin{align}
  \phib \brak{\mathbf{x} + \delta \mathbf{x}
  } =   \phib \brak{\mathbf{x}
  } +  \brak{D \phib}\brak{\mathbf{x}} \cdot \delta \mathbf{x}
  + \mathcal{O}\brak{\|\delta \mathbf{x}\|^{2}} .
\end{align}
Here we see that this derivative takes values in the space of linear operators
on \(\RThree\). More accurately, \(D\phib (\xx) \) is linear mapping between the tangent spaces \( T_{\xx} \BB\)
and \( T_{\phib(\xx)} \RThree\), but this distinction can be ignored because \(\RThree\)'s being flat allows 
tangent vectors based at different points to be trivially identified. Due to its importance within continuum mechanics, 
the derivative of a configuration is denoted by the symbol \(\FF\) and is known as the \textit{deformation gradient}. This definition
extends trivially to a motion, \( \phib:\III \rightarrow \EMB{}\), with the time-\( t \) deformation gradient, \(\FF(t)\), being the derivative of the instantaneous configuration, \(\phib(t)\). The value of the deformation gradient at a particle \(\xx\) and time \(t\) can be written either as \(\FF(t)(\xx)\) or, as will generally be done, \(\FF(\xx,t)\).
Finally, the \textit{Jacobian} of a configuration or motion is then defined as
\begin{align}
  J(\xx,t) = \det\FF(\xx,t)  .
\end{align}
Due to our assumption that \( \phiemb(t) \) has a smooth inverse for each \( t\in\III \), it follows from the inverse function theorem \citep[][p.31]{marsden1994mathematical} that \(\FF(\xx,t)\) takes values in the general linear group \(\mathbf{GL}(3)\) (Appendix \ref{app_sub:groups}). We assume without loss of generality that \( J \) is everywhere positive, meaning that the motion is orientation preserving.

\subsubsection{Variational principle}
\label{sec:var_prin}

To write down a variational principle for the body, we require expressions for its \textit{kinetic} and \textit{potential} energy.
Recall that due to conservation of mass \citep[][p.87, Theorem 5.7]{marsden1994mathematical}, the \textit{referential density}, \(\rho\), can be defined through
\begin{align}
\label{eq:referential_density_defn}
	\rho(\xx) = J(\xx,t)\varrho\braksq{\phib(\xx,t),t}, 
\end{align}
where \(\varrho(\yy,t)\) denotes the density  within the 
instantaneous body, \(\BB_{t}\). The kinetic energy at time \(t\) is  given by
\begin{align}
  \KE(t) = \intvol{\BB}{\frac{1}{2}\rho\brak{\xx}\norm{\vv\brak{\xx,t}}^{2}} .
\end{align}

Meanwhile, we subdivide the body's potential energy into two parts: self-gravitational (\( \PE_{G} \)) and elastic (\( \PE_{E} \)). 
The former is just
\begin{align}
\label{eq:VG_defn}
  \PE_{G}(t) = \frac{1}{2}\intvol{\BB_{t}}{\varrho(\yy,t)\phi(\yy,t)} ,
\end{align}
where the \textit{gravitational potential}, \(\phi\), satisfies the usual \textit{Poisson equation} \citep[e.g][]{landau_lifshitz_2,dahlen1998theoretical}. 
As for the elastic potential energy, we have
\begin{align}
  \PE_{E}(t)= \intvol{\BB}{W\braksq{\xx,\FF(\xx,t)}} ,
\end{align}
where \(W\) denotes the body's \textit{strain-energy function}. The form of \( W \) is constrained by the \textit{principle of material-frame indifference} \citep[e.g.][]{marsden1994mathematical,Truesdell2004}, which requires that
\begin{align}
\label{eq:pmfi_defn}
	W(\xx,\QQ\FF) &= W(\xx,\FF) 
\end{align}
for all rotation matrices \( \QQ \in \so(3) \) and \( \FF \in \mathbf{GL}(3) \) (see Appendix \ref{app_sub:groups}). This implies that the strain-energy depends on \( \FF \) through the \textit{right Cauchy--Green tensor} 
\begin{align}
  \CC = \FF^{T}\FF ,
\end{align}
so that the strain-energy function takes the form
\begin{align}
  W(\xx,\FF) = U(\xx,\CC) .
\end{align}
Following \citet[][`AC16' hereafter]{Al_Attar_2016}, we will include a \textit{stress-glut} representation of a seismic source by allowing the strain-energy function to have the explicit time-dependence
\begin{align}
  W(\xx,t,\FF) = U(\xx,\CC) - \frac{1}{2}\ip{\stressGlut(\xx,t)}{\CC} ,
\end{align}
where the stress glut \( \stressGlut \) is a symmetric second-order tensor field.

All in all, the action describing the elastic body is the integrated difference of the body's kinetic and potential energies:
\begin{align}
\label{eq:elastic_body_basic_action}
  \action 
  &= 
  \inttime{\III}{
            \intvol{\BB}{\braksq{
           		 \frac{1}{2}\rho\norm{\vv}^{2}
           		-W\brak{\xx,t,\FF}
        		-\frac{1}{2}\rho\,\phi\circ\phib
            }}
  } ,
\end{align}
with \( \phi \) understood to satisfy Poisson's equation (defined below, eq.~\ref{eq:poisson}) and where we have pulled \( \PE_{G} \) back to the reference body. Hamilton's principle states that the motion of the body is a stationary point of the action subject to fixed endpoint conditions \citep[\eg][Ch.~7]{marsden1994mathematical}.

\subsubsection{Accounting for gravity variationally}
\label{sec:grav_variational}

There are different ways of incorporating into the variational principle the constraint that \( \phi \) satisfy the Poisson equation. One approach is to solve the Poisson equation explicitly, using Green's functions to express \( \phi \) in terms of the motion and thus allowing the action to be seen as a functional of \( \phib \) alone within a purely referential picture (\eg AC16). Alternatively, one can follow \citet{Woodhouse_2015} and incorporate Poisson's equation into the problem via Lagrange multipliers, thus treating both the motion and the gravitational potential on an equal footing. Within this work we find the latter approach preferable, showing, in contrast to  \citet{Woodhouse_2015}, how it can be implemented within a fully referential framework. 

The Poisson equation is, in weak-form \citep[\eg][]{Chaljub_2004,Maitra_2019},
\begin{align}
\label{eq:poisson}
  \frac{1}{4\pi G}
  \intvol{\RThree}{
    \ip{\nabla\phi}{\nabla\psi}
  }
  +
  \intvol{\BB_{t}}{\varrho\,\psi}
  = 0 ,
\end{align}
where \( \psi \) is an arbitrary test-function and \( G \) Newton's gravitational constant. 
Using the method of Lagrange multipliers, we incorporate the Poisson equation into the variational principle by rewriting action~\eqref{eq:elastic_body_basic_action} as
\begin{align}
  \action 
  &= 
  \inttime{\III}{
            \intvol{\BB}{\braksq{
           		 \frac{1}{2}\rho\norm{\vv}^{2}
           		-W\brak{\xx,t,\FF}
        		-\frac{1}{2}\rho\,\phi\circ\phib
            }}
  }
  -
  \inttime{\III}{\braksq{
    \frac{1}{4\pi G}
    \intvol{\RThree}{
       \ip{\nabla\phi}{\nabla\psi}
    }
    +
    \intvol{\BB}{\rho\,\psi\circ\phib}
  }} ,
\end{align}
with \( \psi \) now viewed as a time-dependent Lagrange multiplier field.  
We then vary the action with respect to \( \phi \) to give
\begin{align}
  \frac{1}{4\pi G}
  \intvol{\RThree}{\ip{\nabla\psi}{\nabla\delta\phi}}
  +
  \frac{1}{2}
  \intvol{\BB}{\rho\,\delta\phi\circ\phib}
  =
  0 ,
\end{align}
which is a weak-form PDE for \( \psi \) whose unique solution is by inspection
\begin{align}
  \psi = \frac{1}{2}\phi .
\end{align}
Substituting this back into the action gives
\begin{align}
\label{eq:elastic_body_basic_action__grav_take_2}
  \action 
  &= 
  \inttime{\III}{
            \intvol{\BB}{\braksq{
           		 \frac{1}{2}\rho\norm{\vv}^{2}
           		-W\brak{\xx,t,\FF}
        		-\rho\,\phi\circ\phib
            }}
  }
  -
  \frac{1}{8 \pi G}
  \inttime{\III}{
    \intvol{\RThree}{
       \mynorm{\nabla\phi}^{2}
    }
  } .
\end{align}
As noted by Woodhouse, variation of this action yields the correct equations of motion.
We have thus eliminated the Lagrange multiplier to give a `reduced' variational principle that has \( \phib \) and \( \phi \) as its independent variables.

Action~\eqref{eq:elastic_body_basic_action__grav_take_2} still mixes the spatial with the referential through its dependence on \( \phi \).
Within the reference body, we can define a \textit{referential potential}, \(\zeta = \phi \circ \phib\) (\eg AC16), but this 
definition does not apply within the exterior domain. Our solution is simply to extend the domain of \( \phib \). Instead of taking \( \phiemb(t) \) to be an embedding of \( \BB \) into \( \RThree \), we take it to be a diffeomorphism that maps 
\(\RThree\) onto itself. For later reference, we note that a diffeomorphism is a smooth mapping with a smooth inverse. The collection 
of such mappings defined on a smooth manifold, \(M\), will be written \(\DIFF{M}\) and forms a group under composition. 
We now see that the instantaneous configurations of the body lie in the diffeomorphism group of Euclidean space, \( \DIFF{} \), while the motion 
is a time-parametrised curve taking values within this group. In Section~\ref{sec:single_body_particle_relabelling} we will show that such an extension of the motion is well-defined.

With the domain of the motion so-extended, we can now define the referential potential globally by 
\begin{subequations}
\label{eq:zeta_defn}
\begin{align}
  &\zeta \equiv \phi\circ\phib
\end{align}
\end{subequations}
along with the tensor \citep[see][]{Maitra_2019}
\begin{align}
  \aa \equiv J \FF^{-1}\FF^{-T} ,
\end{align}
and rewrite action~\eqref{eq:elastic_body_basic_action__grav_take_2} as
\begin{align}
\label{eq:elastic_body_basic_action__grav_take_3}
  \action 
  &= 
  \inttime{\III}{
            \intvol{\BB}{\braksq{
           		 \frac{1}{2}\rho\norm{\vv}^{2}
           		-W\brak{\xx,t,\FF}
        		-\rho\,\zeta
            }}
  }
  -
  \frac{1}{8 \pi G}
  \inttime{\III}{
    \intvol{\RThree}{
       \ip{\nabla\zeta}{\aa\cdot\nabla\zeta}
    }
  } .
\end{align}
This form of the action is central to the rest of the paper. 

\subsection{Some symmetries of the action}
\label{sec:symmetries}

We just extended \( \phib \) arbitrarily from \( \BB \) to the whole of space, a manoeuvre that might raise eyebrows. To see that this is a legitimate step, we need to establish the invariance of the action with respect to a certain transformation group. We will also take this opportunity to discuss other symmetries of the action, which will lead to a discussion of conservation laws and ultimately motivate our introduction of the Tisserand frame.

\subsubsection{Particle-relabelling transformations (or, `Right-actions of the diffeomorphism group')}
\label{sec:single_body_particle_relabelling}
Consider a body with motion \(\phib\) and referential potential \(\zeta\). Given a diffeomorphism, \(\xib\in \DIFF{}\)
we can define new fields by
\begin{align}
  &\brak{\phib,\zeta} 
  \mapsto 
  \brak{\phib\circ\xib,\zeta\circ\xib} .
\end{align}
This mapping defines a \textit{right-action} of the diffeomorphism group, and 
we will denote it by 
\begin{align}
  \brak{\phib,\zeta}\cdot \xib 
  = 
  \brak{\phib\circ\xib,\zeta\circ\xib} .
\end{align}
In the terminology of AC16 and A18, this right-action is known as a \textit{particle-relabelling transformation}. From the kinematic identities discussed above, \( \xib \)'s right-action on \( \phib \) induces transformations of the problem's other variables, for example 
\begin{subequations}
\label{eq:prt_kin_defn}
\begin{align}
  \vv &\mapsto \,\vv\cdot\xib = \vv\circ\xib \\
  \FF &\mapsto \FF\cdot\xib = \brak{\FF\circ\xib}\FF_{\xib} .
  \label{eq:prt_kin_defn__F}
\end{align}
\end{subequations}

Starting from the action in eq.~\eqref{eq:elastic_body_basic_action__grav_take_3}, we can use \(\xib\) to change variables
to obtain
\begin{align}
  \action 
  &= 
  \inttime{\III}{
            \intvol{\tilde{\BB}}{\braksq{
           		 \frac{1}{2}\tilde{\rho}\norm{\tilde{\vv}}^{2}
           		-\tilde{W}\brak{\xx,t,\tilde{\FF}}
        		-\tilde{\rho}\,\tilde{\zeta}
            }}
  }
  -
  \frac{1}{8 \pi G}
  \inttime{\III}{
    \intvol{\RThree}{
       \ip{\nabla\tilde{\zeta}}{\tilde{\aa}\cdot\nabla\tilde{\zeta}}
    }
  } ,
\end{align}
where for convenience we have set \((\tilde{\phib},\tilde{\zeta}) = (\phib,\zeta)\cdot \xib\),  \(\tilde{\rho} = \rho \cdot \xib\), \(\tilde{W} = W \cdot \xib\) and \( \tilde{\BB} = \xib^{-1}\brak{\BB} \), and where we have introduced the right-actions
\begin{subequations}
\label{eq:prt_rho_W_defn}
\begin{align}
  &(\rho\cdot\xib)(\xx)  
  \equiv 
  J_{\xib}(\xx)(\rho\circ\xib)(\xx),
  \label{eq:prt_rho_W_defn__rho}\\
  &\brak{W\cdot\xib}(\xx,t,\FF)
  \equiv 
  J_{\xib}(\xx)W\braksq{\xib(\xx),t,\FF\FF_{\xib}^{-1}} 
  \label{eq:prt_rho_W_defn__W}
\end{align}
\end{subequations}
for arbitrary \((\xx,t) \in \RThree\times \III\)  and \( \FF\in\gl(3) \). This argument has not, in general, demonstrated an invariance 
of the action because it has been necessary to transform the material parameters \((\rho,W)\); we will return later to the question of whether
these parameters can be left unchanged by the action of a non-trivial diffeomorphism. Nevertheless, if we restrict \(\xib\) such that 
it equals the identity on an open subset containing \(\BB\) then the material parameters \textit{are} unaltered and 
we have identified a symmetry of the action. Denoting by \(\DIFFRest{ \BB}{\RThree}\) the subgroup of such diffeomorphisms, 
we conclude that if \((\phib,\zeta)\) is a solution of the equations of motion, then so is \((\phib,\zeta)\cdot \xib\) 
for any \(\xib\in \DIFFRest{\BB}{\RThree}\). These two solutions agree within the reference body, while in 
the exterior domain they are such that the spatial potential \(\phi = \zeta \circ \phib^{-1}\) is  uniquely defined.
Here we see an example a \textit{gauge invariance}, with the solution of the problem only being defined up to the action 
of a certain transformation group, but with all physically meaningful quantities being unaffected by such transformations. We can, in solving the problem,
fix these additional degrees of freedom in any manner that is convenient.  In more 
formal terms, we have extended the configuration space of the body to \(\DIFF{}\), but have shown that  the dynamics
is defined only up to an element of \(\DIFFRest{\BB}{\RThree}\). This means that a unique solution can be found within the quotient space
\(\DIFF{} / \DIFFRest{\BB}{\RThree}\) which precisely equals the initial configuration space \(\EMB{}\).

\subsubsection{Left-actions of the diffeomorphism group}
\label{sec:left_actions}

One can also carry out transformations where the diffeomorphism group acts from the left. Similarly to right-actions, such \textit{left-actions} of \( \DIFF{} \) are not generally a symmetry of the action. But we do have a symmetry when acting with the \textit{Euclidean group} \( \EThree \), a subgroup of \( \DIFF{} \). 
That symmetry underlies the rest of this paper.

Consider some \( \phiemb \in \DIFF{} \) and let \( \Diffel \in \DIFF{} \). By the diffeomorphism group's definition \( \Diffel\circ\phiemb \) is also an element of \( \DIFF{} \), so we can define a left-action of \( \DIFF{} \) on itself by
\begin{align}
\label{eq:diff_left_action_phib}
  \phiemb \mapsto \Diffel\cdot\phiemb \equiv \Diffel\circ\phiemb .
\end{align}
The corresponding left-action of \( \Diffel \) on the \textit{spatial} potential \( \phi \) is
\begin{align}
  \phi \mapsto \Diffel\cdot\phi = \phi\circ\brak{\Diffel^{-1}},
\end{align}
so given eqs.~\eqref{eq:zeta_defn} and \eqref{eq:diff_left_action_phib}, \( \Diffel \)'s action on \( \zeta \) is trivial:
\begin{align}
  \zeta \mapsto \Diffel\cdot\zeta = \zeta .
\end{align}

To be able to transform action~\eqref{eq:elastic_body_basic_action__grav_take_3} using such left-actions we must also consider how derivatives of \( \phib \) are transformed. Therefore consider some time-dependent embedding (\ie a motion) \( \phiemb : \III \mapsto \DIFF{} \). The associated tangent-vector at \( t=0 \) is
\begin{align}
  \evaluateat{\dbyd{t}\phiemb(t)}{t=0}
  \in
  \Tang{\phiemb(0)}{\DIFF{}} .
\end{align}
Given \( \Diffel \in \DIFF{} \) we can then consider
\begin{align}
  t 
  &\mapsto 
  \brak{\Diffel\cdot\phiemb}(t)
  \equiv 
  \brak{\Diffel\circ\phiemb}(t)
  = 
  \Diffel\braksq{\phiemb(t)} ,
\end{align}
with associated tangent vector
\begin{align}
  \evaluateat{\dbyd{t}\brak{\Diffel\cdot\phiemb}(t)}{t=0}
  =
  \braksq{D\Diffel\circ\phiemb(0)}
  \evaluateat{\dbyd{t}\phiemb(t)}{t=0}
  \in 
  \Tang{\Diffel\cdot\phiemb(0)}{\DIFF{}} .
\end{align}
Thus we see that \( \Diffel \) induces a linear mapping from \( \Tang{\phiemb(0)}{\DIFF{}} \) to \( \Tang{\Diffel\cdot\phiemb(0)}{\DIFF{}} \). This is of course true for any \( t \), so for some \( \phiemb\in\DIFF{} \) and \( \vemb \in \Tang{\phiemb}{\DIFF{}} \) we set
\begin{align}
\label{eq:diff_left_action_vv}
  \Diffel\cdot\vemb = \brak{D\Diffel\circ\phiemb}\cdot\vemb ,
\end{align}
which defines the left-action of \( \DIFF{} \) on \( \Tang{}{\DIFF{}} \). One may also show readily that the deformation gradient transforms to
\begin{align}
\label{eq:diff_left_action_FF}
  \Diffel\cdot\FF = \brak{D\Diffel\circ\phiemb}\FF ,
\end{align}
from which one may easily find \( \DIFF{} \)'s action on the referential Poisson tensor \( \aa \mapsto \Diffel\cdot\aa \).

The action on \( \phib \) of \( \DIFF{} \)'s Lie algebra \( \diff{} \) is also important in what follows.
We need only focus on the motion \( \phib \) because \( \zeta \) is left-invariant. To proceed, fix \( \phiemb \in \DIFF{} \) and consider
\begin{align}
  s \mapsto \Diffel(s) \in \DIFF{}
\end{align}
where \( \Diffel(0) = \idgen \) (and with \( \idgen \) denoting the identity mapping on a set). We can define
\begin{align}
  \dot{\Diffel}(0)
  \equiv 
  \evaluateat{\dbyd{s}\Diffel(s)}{s=0}
  \in 
  \Tang{\Diffel(0)}{\DIFF{}},
\end{align}
and since \( \Diffel(0) = \idgen \) the RHS is by definition \( \DIFF{} \)'s Lie algebra, \( \diff{} \).
We can now form the curve in \( \DIFF{} \)
\begin{align}
  s \mapsto \Diffel(s)\cdot\phiemb = \Diffel(s)\circ\phiemb ,
\end{align}
whose tangent vector at the origin is
\begin{align}
  \evaluateat{\dbyd{s}\Diffel(s)\cdot\phiemb}{s=0}
  =
  \dot{\Diffel}(0)\circ\phiemb .
\end{align}
Given that \( \Diffel(0) = \idgen \), the LHS is an element of \( \Tang{\phiemb}{\DIFF{}} \)\ while the RHS represents the action of \( \diff{} \) on \( \DIFF{} \). We have obtained a mapping of \( \DIFF{} \) into \( \Tang{\phiemb}{\DIFF{}} \) by \( \diff{} \), which we write as
\begin{align}
  \dot{\Diffel}\cdot\phiemb 
  =
  \dot{\Diffel}\circ\phiemb .
\end{align}
This is the desired left-action of \( \diff \).

All in all, action~\eqref{eq:elastic_body_basic_action__grav_take_3} transforms under \( \DIFF{} \)'s left-action into
\begin{align}
\label{eq:diff_group_action_not_invariant}
  \inttime{\III}{
            \intvol{\BB}{\braksq{
           		 \frac{1}{2}\rho\norm{\brak{D\Diffel\circ\phiemb}\cdot\vv}^{2}
           		-W\brak{\xx,t,\brak{D\Diffel\circ\phiemb}\FF}
        		-\rho\,\zeta
            }}
  }
  -
  \frac{1}{8 \pi G}
  \inttime{\III}{
    \intvol{\RThree}{
      \ip{\nabla\zeta}{\brak{\Diffel\cdot\aa}\cdot\nabla\zeta} 
    }
  } .
\end{align}
For general \( \Diffel\in\DIFF{} \) it is clear that this does not equal the original action; hence the action is not left-invariant under the whole of \( \DIFF{} \). It would be invariant if we could somehow `remove' the terms in \( D\Diffel \) -- but to do that we need to restrict attention to a certain subgroup of \( \DIFF{} \): the \textit{Euclidean group} \( \EThree \).

\subsubsection{The Euclidean group and its left-actions}
\label{sec:left_actions_euc}

The elements of \( \EThree \subseteq \DIFF{} \) are the \textit{isometries} of 3D space: translations and rotations. \( \EThree \) is the \textit{semi-direct product} of \( \TThree \) and \( \SOThree \), the groups of translations and rotations respectively, which are discussed in Appendix~\ref{app_sub:groups}. We write an arbitrary element \( \Euc \in \EThree \) as
\begin{align}
  \Euc = \EucComp{\Phib}{\RR}
\end{align}
for some \( \Phib\in\RThree \) and \( \RR\in\SOThree \); \( \Euc \)'s action on a motion \( \phiemb \in \DIFF{} \) is (\cf eq.~\ref{eq:diff_left_action_phib})
\begin{align}
\label{eq:euc_action_motion}
  \Euc\cdot\phiemb = \Phib + \RR\cdot\phiemb,
\end{align}
simultaneously translating \( \phiemb \) by \( \Phib \) and rotating it by \( \RR \). The inverse transformation is
\begin{align}
  \Euc^{-1} = \EucComp{-\RR^{T}\cdot\Phib}{\RR^{T}} .
\end{align}

Correspondingly, the left-action of the Euclidean group's Lie algebra on the motion is given by
\begin{align}
  \phiemb 
  &
  \mapsto 
  \evaluateat{\dbyd{s}\Euc(s)\cdot\phiemb}{s=0} ,
\end{align}
for some curve \( s \mapsto \Euc(s) \) with \( \Euc(0) \) the identity. 
We show in Appendix~\ref{sec:groups_Tthree} that the translation group's Lie algebra consists of all velocities \( \eucv\in\RThree \), and that \( \trans\in\tthree \) acts like
\begin{align}
  \trans\cdot\phib = \eucv .
\end{align} 
On the other hand, \( \SOThree \)'s Lie algebra \( \sothree \) contains all possible antisymmetric matrices \( \eucom \) (Appendix~\ref{sec:groups_sothree}). In the context of rigid body mechanics these are known as angular velocities, acting on \( \phib \) like 
\( \eucom\cdot\phib \).
It follows readily that \( \ethree \)'s action is
\( 
    \phib
    \mapsto
    \eucv + \eucom\cdot\phiemb 
\).
We therefore define an element \( \euc\in\ethree \) to be
\begin{align}
  \euc    &=   \eucComp{\eucv}{\eucom},
\end{align}
with its action on a motion described by
\begin{align}
\label{eq:euc_lie_action_motion}
  \euc\cdot\phib = \eucv + \eucom\cdot\phib .
\end{align}

Recall from Appendix \ref{app_sub:groups} (eq.~\ref{eq:rigid_body:big_ad_defn}) that the rotation group's adjoint representation \( \BigAd{} \) allows \( \SOThree \) to rotate elements of \( \sothree \) into other elements of \( \soThree \). 
Analogously, we may define the adjoint representation of the Euclidean group, \( \BigAd{}: \ethree\times\EThree \mapsto \ethree \), whose action is
\begin{align}
\label{eq:big_Ad_defn}
  \BigAd{\Euc}\cdot\euc
  =
  \Euc\euc\Euc^{-1} .
\end{align}
The action of \( \BigAd{\Euc} \) on \( \euc = \eucComp{\eucv}{\eucom} \) produces a new element of \( \ethree \) that acts on an arbitrary motion as
\begin{align}
  \brak{\BigAd{\Euc}\cdot\euc}\cdot\phib 
  &=
  \Euc\cdot\braksq{\eucv + \eucom\cdot\brak{\Euc^{-1}\cdot\phib}}
  \nonumber\\
  &=
  \RR\cdot\braksq{\eucv + \eucom\RR^{-1}\cdot\brak{\phib-\Phib}}
  \nonumber\\
  &=
  \RR\cdot\eucv 
  + 
  \brak{\BigAd{\RR}\cdot\eucom}\cdot\brak{\phib-\Phib} .
\end{align}

With these preliminaries we can return to \( \EThree \)'s left-actions on \( \action \). We see immediately from eq.~\eqref{eq:euc_action_motion} that \( D\Euc\circ\phiemb = \RR \in \SOThree \), so by restricting eqs.~\eqref{eq:diff_left_action_phib}, \eqref{eq:diff_left_action_vv} and \eqref{eq:diff_left_action_FF} from \( \DIFF{} \) to \( \EThree \), we find that
\begin{subequations}
\begin{align}
  \phib &\mapsto \Euc\cdot\phib = \Phib + \RR\cdot\phib
  \\
  \vv   &\mapsto \Euc\cdot\vv   = \RR\cdot\vv
  \\
  \FF   &\mapsto \Euc\cdot\FF   = \RR\FF .
\end{align}
\end{subequations}
Crucially, 
looking back to expression~\eqref{eq:diff_group_action_not_invariant} and recalling the Principle of Material Frame Indifference (eq.~\ref{eq:pmfi_defn}), it follows that
\textit{the action is invariant under the left-action of the whole Euclidean group.}

\subsection{Noether's theorem}
\label{sec:solid_noether}

Now that we have established various symmetries of the action, we can use Noether's theorem to study their associated conservation laws. We start by deriving the general result in the appropriate context, then we apply it to obtain some familiar results. Later (Section~\ref{sec:vorticity_maybe}) we show that the same ideas can be applied within fluid regions to obtain an apparently novel conservation law.

Consider an action taking the general form
\begin{align}
  \action\braksq{\phib}
  =
  \inttime{\III}{\intvol{\BB}{
    \lag\brak{\phib,\vv,\FF}
  }} ,
\end{align}
where on the LHS we make explicit the action's functional dependence on the motion. 
Its variation is
\begin{align}
  \delta\action
  =
  \inttime{\III}{\intvol{\BB}{\brak{
     \ip{D_{\phib}\lag}{\delta\phib}
    +\ip{D_{\vv}\lag}{\delta\vv}
    +\ip{D_{\FF}\lag}{\delta\FF}
  }}} ,
\end{align}
and, given that 
\begin{align}
  \ip{D_{\vv}\lag}{\delta\vv}
  =
  \partial_{t}\ip{D_{\vv}\lag}{\delta\phib}
  -
  \ip{\partial_{t}\brak{D_{\vv}\lag}}{\delta\phib} ,
\end{align}
the variation satisfies
\begin{align}
\label{eq:noether_variation}
  \delta\action
  =
  \inttime{\III}{\intvol{\BB}{\brak{
     \ip{
      D_{\phib}\lag
      -
      \partial_{t}\brak{D_{\vv}\lag}
     }{
      \delta\phib
     }
    +\ip{D_{\FF}\lag}{\delta\FF}
  }}} 
  +
  \braksq{
    \intvol{\BB}{\ip{D_{\vv}\lag}{\delta\phib}}
  }_{t_{0}}^{t_{1}} .
\end{align}
If the variation \( \delta\phib \) vanishes at the start- and endpoints, then we may appeal to Hamilton's principle and demand that \( \delta\action = 0 \), leading to the weak-form Euler--Lagrange equations
\begin{align}
\label{eq:noether_on_shell}
  \intvol{\BB}{
     \ip{
      D_{\phib}\lag
      -
      \partial_{t}\brak{D_{\vv}\lag}
     }{
      \delta\phib
     }
  }
  +
  \intvol{\BB}{
     \ip{D_{\FF}\lag}{\delta\FF}
  } 
  =
  0
\end{align}
for time-independent test functions \( \delta\phib \) and with \( \delta\FF = D\brak{\delta\phib} \). 

Now consider the one-parameter family of fields
\begin{align}
  s \mapsto \phib_{s} ,
\end{align}
where \( \phib_{0} \) is a particular solution to eq.~\eqref{eq:noether_on_shell}; we describe \( \phib_{0} \) as `on-shell'. 
If the condition
\begin{align}
\label{eq:noether_symmetry_condition}
  \action\braksq{\phib_{s}} = \action\braksq{\phib_{0}}
\end{align}
is satisfied for all \( s \), then the family \( \phib_{s} \) is a \textit{one-parameter continuous symmetry} of the action. Differentiating this with respect to \( s \) about \( s = 0 \) gives
\begin{align}
\label{eq:noether_symmetry_condition_deriv}
  \evaluateat{
    \dbyd{s}\action\braksq{\phib_{s}}
  }{s=0}
  = 0 ,
\end{align}
which is sometimes more practically useful.
This derivative is given explicitly by eq.~\eqref{eq:noether_variation}, but with
\begin{align}
  \delta\phib = \evaluateat{\dbyd{s}\phib_{s}}{s=0} .
\end{align}
This \( \delta\phib \) need not satisfy the vanishing-endpoint condition, but \( \phib_{0} \) itself is on-shell, so eq.~\eqref{eq:noether_symmetry_condition_deriv} reduces to
\begin{align}
\label{eq:noether_tmp_654794641}
  \braksq{
    \intvol{\BB}{
      \ip{
        D_{\vv}\lag
      }{
        \evaluateat{\dbyd{s}\phib_{s}}{s=0}
      }
    }
  }_{t_{0}}^{t_{1}}
  =
  0 .
\end{align}
Given that the start- and end-times are arbitrary it follows that the quantity
\begin{align}
\label{eq:noether_theorem}
  Q 
  \equiv 
  \intvol{\BB}{
    \ip{
      D_{\vv}\lag
    }{
      \evaluateat{\dbyd{s}\phib_{s}}{s=0}
    }
  }
\end{align}
is conserved in time. 
This is Noether's theorem. The result shows concretely how the conserved quantity corresponds to the symmetry generating it, and gives a means of computing it.

\subsubsection{(Angular-) momentum conservation}
\label{sec:momentum_cons}

For the specific action~\eqref{eq:elastic_body_basic_action__grav_take_3} we now consider a definite solution \( \phib_{0} \) and \( \zeta_{0} \). Then we define the curve \( s \mapsto \Euc(s) \in \EThree \),
and thus define the one-parameter family
\begin{align}
  s 
  \mapsto \brak{\phib_{s},\zeta_{s}} 
  = \Euc(s)\cdot\brak{\phib_{0},\zeta_{0}}
\end{align}
with \( \Euc(0) \) equal to the identity.
Given that the action is invariant under the left-action of \( \EThree \), \( \phib_{s} \) and \( \zeta_{s} \) satisfy
\begin{align}
  \action\braksq{\phib_{s},\zeta_{s}}
  =
  \action\braksq{\phib_{0},\zeta_{0}}
\end{align}
and are therefore a one-parameter symmetry of the action (\cf  eq.~\ref{eq:noether_symmetry_condition}). Since
\begin{align}
  \evaluateat{\dbyd{s}\phib_{s}}{s=0}
  =
  \brak{
    \evaluateat{\dbyd{s}\Euc(s)}{s=0}
  }
  \cdot\phib
  \equiv
  \euc\cdot\phib ,
\end{align} 
Noether's theorem (eq.~\ref{eq:noether_theorem}) shows that
\begin{align}
  \dbyd{t}\intvol{\BB}{\rho\ip{\vv}{\euc\cdot\phib}} = 0 .
\end{align}
\( \euc \) is arbitrary, so we have shown the existence of  
a conserved quantity \( \pib \in \ethree \), which we define implicitly through
\begin{align}
\label{eq:mom_defn}
  \fip{\pib}{\euc} = \intvol{\BB}{\rho\ip{\vv}{\euc\cdot\phib}} 
\end{align}
for arbitrary \( \euc \in \ethree \) and where the \(\ethree\)-inner-product on the LHS is defined in the obvious manner.
We refer to \( \pib \) as the \textit{generalised momentum}. From eq.~\eqref{eq:euc_lie_action_motion}
\begin{align}
  \fip{\pib}{\euc}
  =
  \intvol{\BB}{\rho\ip{\vv}{\eucv}}
  +
  \intvol{\BB}{\rho\ip{\vv}{\eucom\cdot\phib}} ,
\end{align}
and we may use the \textit{wedge product} defined by eq.~\eqref{eq:rigid_body_wedge_defn} to rearrange this as
\begin{align}
  \fip{\pib}{\euc}
  =
  \ip{\eucv}{\intvol{\BB}{\rho\vv}}
  +
  \ip{\eucom}{\intvol{\BB}{\rho\vv\wedge\phib}} .
\end{align}
Since \( \eucv \) and \( \eucom \) are arbitrary, \( \intvol{\BB}{\rho\vv} \) and \( \intvol{\BB}{\rho\vv\wedge\phib} \) are both conserved. Conservation of the generalised momentum \( \pib \) is thus seen to encompass conservation of both the linear and angular momenta of the elastic body. For later reference we define
\begin{subequations}
\label{eq:conserved_momenta_defns_initial}
\begin{align}
  \mom &\equiv \intvol{\BB}{\rho\vv}
  \\
  \angmom &\equiv \intvol{\BB}{\rho\vv\wedge\phib} .
\end{align}
\end{subequations}
We remark that this exercise has also demonstrated that the wedge product is analogous to the standard cross product.

\subsubsection{Energy conservation}

If we temporarily neglect the strain-energy's stress-glut term, then action~\eqref{eq:elastic_body_basic_action__grav_take_3} is also invariant when its fields are shifted by a time \( \tau \) and a corresponding shift is made in \( \III \). That is, if we define
\begin{align}
  &\phib_{\tau}(t) = \phib_{0}(t+\tau)
  \\
  &\,\zeta_{\tau}(t)\, = \zeta_{0}(t+\tau)
  \\
  &\III_{\tau} = \brakbr{t-\tau \,\,\vline\,\, t\in\III_{0}} ,
\end{align} 
then
\begin{align}
  \action\braksq{\phib_{\tau},\zeta_{\tau},\III_{\tau}}
  =
  \action\braksq{\phib_{0},\zeta_{0},\III_{0}}
  ,
\end{align}
where we are temporarily viewing the action as a function of the time-interval as well. The one-parameter family
\begin{align}
  \tau \mapsto \brak{\phib_{\tau},\zeta_{\tau},\III_{\tau}}
\end{align}
is therefore another one-parameter continuous symmetry of the action.
It follows that
\begin{align}
  \evaluateat{\dbyd{\tau}\action\braksq{\phib_{\tau},\zeta_{\tau},\III_{\tau}}}{\tau=0}
  =
  \braksq{
    \intvol{\BB}{
      \ip{
        D_{\vv}\lag
      }{
        \evaluateat{\dbyd{\tau}\phib_{\tau}}{\tau=0}
      }
    }
  }_{t_{0}}^{t_{1}}
  -
  \braksq{
    \intvol{\BB}{
      \lag 
    }
  }_{t_{0}}^{t_{1}}
  =
  0 ,
\end{align}
with the term in \( \lag \) 
arising from the variation of \( \III \). There is no term here in \( \zeta_{\tau} \) because the action does not depend on \( \zeta \)'s time-derivatives. Concretely, Noether's theorem thus implies that the total energy
\begin{align}
  \intvol{\BB}{\braksq{
 	 \frac{1}{2}\rho\norm{\vv}^{2}
 	+W\brak{\xx,\FF}
	+\rho\zeta
  }}
  +
  \frac{1}{8 \pi G}
  \intvol{\RThree}{
     \ip{\nabla\zeta}{\aa\cdot\nabla\zeta}
  }
\end{align}
is conserved.

\subsection{Decomposition of the motion}
\label{sec:decomposition_solid}

Motivated by the action's invariance under a constant Euclidean motion, we now introduce the Tisserand frame by writing \( \phib \) as an elastic motion upon which a \textit{time-dependent} Euclidean motion is imposed:
\begin{align}
\label{eq:decomposition}
  \phiemb(t) = \Euc(t)\cdot\phiembr(t) ,
\end{align}
where \( \Euc : \III \mapsto \EThree \) is the \textit{Euclidean motion}, which captures the motion of the frame, and where \( \phiembr : \III \mapsto \DIFF{} \) is what we term the \textit{internal motion}. Now, 
\( \Euc(t) \)'s left-action will alter the action via \( \KE \), so this might seem to complicate the problem. However, introducing the new function \( \Euc(t) \) frees us -- indeed, obliges us -- to impose constraints upon \( \phibr \), and we will choose those constraints so that \( \phibr \) has no secular behaviour and couples minimally to the Euclidean motion. 
The remainder of this section lays all this out. In what follows we will generally suppress the respective motions' time-arguments.

\subsubsection{The decomposition's kinematics}

Given eq.~\eqref{eq:decomposition}, the corresponding velocity is
\begin{align}
  \vemb = \Euc\cdot\vembr + \dot{\Euc}\cdot\phiembr ,
\end{align}
with \( \vvr(\xx,t) \equiv \dot{\phibr}(\xx,t) \).
To write the velocity more conveniently we define
\begin{subequations}
\label{eq:ur_reconstruction_equation}
\begin{align}
  \euc  &\equiv \dot{\Euc}\Euc^{-1} 
  \label{eq:ur_reconstruction_equation_spatial} 
  \\
  \eucr &\equiv \Euc^{-1}\dot{\Euc}
  \label{eq:ur_reconstruction_equation_internal} ,
\end{align}
\end{subequations}
both of which take values in \( \ethree \). \( \euc \) and \( \eucr \) are related by \( \Euc \)'s adjoint representation (eq.~\ref{eq:big_Ad_defn}):
\begin{align}
  \euc = \BigAd{\Euc}\cdot\eucr .
\end{align}
It now follows that \( \vv \) may be written in two distinct ways:
\begin{subequations}
\label{eq:velocity_diff}
\begin{align}
  \vemb 
  &= \Euc\cdot\vembr + \euc\cdot\phiemb
  \label{eq:velocity_diff__1}\\
  &= \Euc\cdot\brak{\vembr + \eucr\cdot\phiembr} 
  \label{eq:velocity_diff__2} .
\end{align}
\end{subequations}
These expressions show that one may choose to express the velocity such that it features either the internal motion \( \phiembr \) or the original motion \( \phiemb \). 

Similarly to the velocity, we may write the conserved generalised momentum \( \pib \) (eq.~\ref{eq:mom_defn}) in two different ways. Using eqs.~(\ref{eq:decomposition}, \ref{eq:velocity_diff__1}), 
\begin{align}
\label{eq:euclidean_conserved_qty_spatial_0}
  \fip{\pib}{\kapb}
  &=
  \intvol{\BB}{\rho\ip{\Euc\cdot\vvr}{\kapb\cdot\phib}}
  +
  \intvol{\BB}{\rho\ip{\euc\cdot\phib}{\kapb\cdot\phib}},
\end{align}
where \(\kapb\) denotes an arbitrary element of \( \ethree \).
We now define the spatial and internal \textit{effective mass tensors} 
\( \JJ, \JJr \) 
to be
\begin{subequations}
\label{eq:effective_mass_defn}
\begin{align}
  \fip{\JJ\cdot\kapb'}{\kapb}    &\equiv \intvol{\BB}{\rho\ip{\kapb'\cdot\phib}{\kapb\cdot\phib}} 
  \label{eq:effective_mass_defn__spatial}\\
  \fip{\JJr\cdot\kapb'}{\kapb} &\equiv \intvol{\BB}{\rho\ip{\kapb'\cdot\phibr}{\kapb\cdot\phibr}} 
  \label{eq:effective_mass_defn__internal} 
\end{align}
\end{subequations}
with \( \kapb' \in \ethree \) also arbitrary.
They map \( \ethree \) onto itself; they are clearly symmetric and positive-definite, and therefore invertible.
If we then set
\begin{align}
\label{eq:kappa_bar_defn}
  \kapbr \equiv \BigAd{\Euc^{-1}}\cdot\kapb ,
\end{align}
the second term on eq.~\eqref{eq:euclidean_conserved_qty_spatial_0}'s RHS may be written either as \( \fip{\JJ\cdot\euc}{\kapb} \) or \( \fip{\JJr\cdot\eucr}{\kapbr} \).
Next, we use \( \kapbr \)'s definition to write
\begin{align}
  \intvol{\BB}{\rho\ip{\Euc\cdot\vvr}{\kapb\cdot\phib}}
  &=
  \intvol{\BB}{\rho\ip{\vvr}{\kapbr\cdot\phibr}} .
\end{align}
After defining the \textit{internal momentum} \( \pibr \) by
\begin{align}
\label{eq:pi_internal_defn}
  \fip{\pibr}{\kapb} \equiv \intvol{\BB}{\rho\ip{\vvr}{\kapb\cdot\phibr}} ,
\end{align}
the momentum \( \pib \) may finally be written as:
\begin{subequations}
\label{eq:pi_internal_spatial}
\begin{align}
    \pib
    &= \BigAd{\Euc}\cdot\pibr + \JJ\cdot\euc
    \label{eq:pi_internal_spatial__1} \\
    &= \BigAd{\Euc}\cdot\brak{\pibr+\JJr\cdot\eucr} .
    \label{eq:pi_internal_spatial__2}
\end{align}
\end{subequations}
Eq.~\eqref{eq:pi_internal_spatial__1} is the more useful for Section~\ref{eq:solid_decomp_uniqueness} while \eqref{eq:pi_internal_spatial__2} will be used almost exclusively from Section~\ref{sec:var_prin_solid_decomp} onwards.

\subsubsection{The decomposition's uniqueness}
\label{eq:solid_decomp_uniqueness}

As written, decomposition~\eqref{eq:decomposition} is non-unique. That is, given a definite \( \phib \) one cannot uniquely construct \( \Euc \) and \( \phibr \). This is for two similar but distinct reasons. Firstly, we have introduced six extra degrees of freedom at each time \( t \) through the new function \( \Euc \). Secondly, the decomposition is only defined up to an arbitrary Euclidean motion: if eq.~\eqref{eq:decomposition} holds for some \( \Euc \) and \( \phibr \), then it also holds for \( \Euc\tilde{\Euc}^{-1} \) and \( \tilde{\Euc}\cdot\phibr \) with \( \tilde{\Euc}\in\EThree \) arbitrary. However, requiring the internal motion's velocity not to contribute to the internal momentum is sufficient to make decomposition~\eqref{eq:decomposition} unique up to initial conditions on \( \Euc \), and we can in fact choose those to be whatever is convenient.

To begin, consider eq.~\eqref{eq:pi_internal_spatial__1}'s statement that the conserved momentum \( \pib \) is equal to the internal momentum \( \pibr \) plus a term related to \( \euc \) and \( \phib \). \( \pibr \) represents the linear and angular momentum associated with the internal elastic motion \( \phibr \), so in order to 
satisfy the Tisserand condition
it seems desirable to demand that \( \pibr \) vanish. 
We therefore set
\begin{align}
\label{eq:constraint}
  \pibr(t) = 0 \quad \forall t\in\III .
\end{align}
to be a constraint associated with decomposition~\eqref{eq:decomposition}.

To prove that this constraint is sound, we start by noting that it reduces eq.~\eqref{eq:pi_internal_spatial__1} to the statement
\begin{align}
\label{eq:gen_mom_JJ_euc}
  \pib = \JJ\cdot\euc .
\end{align}
Using \( \JJ \)'s invertibility and \( \euc \)'s definition~\eqref{eq:ur_reconstruction_equation_spatial}, this implies that
\begin{align}
\label{eq:recons_tmp}
  \dot{\Euc} = \brak{\JJ^{-1}\cdot\pib}\Euc .
\end{align} 
Given a definite motion \( \phib \) we can directly construct both \(\pib\) and \(\JJ\), then solve this ODE to give \( \Euc \). The final step is to write
\begin{align}
\label{eq:recons_tmp_2}
  \phiembr(t) = \Euc(t)^{-1}\cdot\phiemb(t) ,
\end{align}
inverting eq.~\eqref{eq:decomposition} to find \( \phiembr \). The solution to eq.~\eqref{eq:recons_tmp} is unique only up to an arbitrary initial condition \( \Euc_{0} \), and therefore so is expression~\eqref{eq:recons_tmp_2} for \( \phibr \). However, an arbitrary change of initial condition \( \Euc_{0} \rightarrow \Euc_{0}\tilde{\Euc} \) merely transforms \( \Euc \) and \( \phibr \) by
\begin{align}
\label{eq:euclidean_gauge_freedom}
  \brak{\Euc, \phiembr} \rightarrow \brak{\Euc\tilde{\Euc}^{-1},\tilde{\Euc}\cdot\phiembr} .
\end{align}
Such a transformation reflects the decomposition's inherent non-uniqueness, mentioned just above. Crucially, it has no effect on the physically observable motion \( \phiemb \). The arbitrary initial condition \( \Euc_{0} \) is thus a further example of a \textit{gauge freedom}. Therefore we may reasonably describe eqs.~\eqref{eq:decomposition}~and~\eqref{eq:constraint} as together constituting a unique decomposition of the motion. 


\subsubsection{A pragmatic notation and choice of initial conditions}
\label{sec:pragmatism}


It will now be convenient to reintroduce an explicit separation between the translational and rotational components of the Euclidean motion, writing
\begin{subequations}
\begin{align}
  \Euc(t) &= \EucComp{\Phib(t)}{\RR(t)},
\end{align}
\end{subequations} 
with \( \Phib : \III \rightarrow \RThree \) and \( \RR : \III \rightarrow \SOThree \).
The motion is then written as
\begin{align}
\label{eq:decomp_explicit__phi}
  \phib = \Phib + \RR\cdot\phibr .
\end{align}
To write the velocity we need the actions of \( \euc \) and \( \eucr \). This leads us to define the \textit{spatial angular velocity} and \textit{body angular velocity} by
\begin{subequations}
\begin{align}
  \Omb  &\equiv \dot{\RR}\RR^{T}
  \\
  \Ombr &\equiv \RR^{T}\dot{\RR} ;
\end{align}
\end{subequations}
both take values in \( \soThree \).
Then \( \euc \)'s action is (eq.~\ref{eq:ur_reconstruction_equation_spatial})
\begin{align}
  \euc\cdot\phib 
  &=
  \EucComp{\VV}{\dot{\RR}}
  \cdot
  \brak{
    \EucComp{-\RR^{T}\cdot\Phib}{\RR^{T}}
    \cdot
    \phib
  }
  \nonumber\\
  &=
  \VV
  -
  \Omb 
  \cdot
  \brak{\Phib - \phib} ,
\end{align}
while for \( \eucr \) (eq.~\ref{eq:ur_reconstruction_equation_internal}; recall that \( \EThree \)'s elements simply rotate velocities)
\begin{align}
\label{eq:eucr_action}
  \eucr\cdot\phibr
  &=
  \EucComp{-\RR^{T}\cdot\Phib}{\RR^{T}}
  \cdot
  \brak{\EucComp{\VV}{\dot{\RR}}\cdot\phibr}
  \nonumber\\
  &=
  \RR^{T}\cdot\VV
  +
  \Ombr\cdot\phibr .
\end{align}
Substituting these explicit actions of \( \ethree \) into eqs.~\eqref{eq:velocity_diff} shows that the velocity is written as
\begin{subequations}
\label{eq:velocity_diff_explicit}
\begin{align}
  \vv  
  &=
  \VV
  -
  \Omb\cdot\Phib
  +
  \RR\cdot\vvr 
  +
  \Omb\cdot\phib
  \label{eq:velocity_diff_explicit__1}\\
  &=
  \VV
  +
  \RR
  \cdot 
  \brak{\vvr+\Ombr\cdot\phibr}
  \label{eq:velocity_diff_explicit__2} .
\end{align}
\end{subequations}

Next, to consider the generalised momentum we should start by introducing the \textit{spatial}  and \textit{internal} \textit{moment of inertia (MoI)} tensors:
\begin{subequations}
\label{eq:moi_defn}
\begin{align}
  \II\cdot\Omb'
  &\equiv 
  \intvol{\BB}{\rho\braksq{\Omb'\cdot\brak{\phib-\Phib}}\wedge\brak{\phib-\Phib}}
  \label{eq:moi_defn_spat}
  \\
  \IIr\cdot\Omb'
  &\equiv 
  \intvol{\BB}{\rho\brak{\Omb'\cdot\phibr}\wedge\phibr}
  \label{eq:moi_defn_int}
\end{align}
\end{subequations}
for arbitrary \( \Omb'\in\sothree  \). We also note that the constraint \( \pibr = 0 \) is equivalent to conservation of internal linear and angular momentum:
\begin{subequations}
\label{eq:constraint_explicit_tmp}
\begin{align}
  &\intvol{\BB}{\rho\vvr} = 0
  \label{eq:constraint_explicit_tmp__P}\\
  &\intvol{\BB}{\rho\vvr\wedge\phibr} = 0 
  \label{eq:constraint_explicit_tmp__J}.
\end{align}
\end{subequations}
It then follows upon substitution of eqs.~(\ref{eq:decomp_explicit__phi},~\ref{eq:velocity_diff_explicit}) into eqs.~\eqref{eq:conserved_momenta_defns_initial} that the conserved spatial linear and angular momenta take the forms
\begin{subequations}
\label{eq:total_momentum}
\begin{align}
  \mom
  &=
  m\VV
  +
  \Omb\RR\cdot\intvol{\BB}{\rho\phibr}
  \label{eq:total_momentum__linear}\\
  \angmom 
  &=
  m\VV\wedge\Phib
  +
  \II\cdot\Omb
  +
  \brak{\Omb\RR\cdot\intvol{\BB}{\rho\phibr}}\wedge\Phib
  +
  \vvr\wedge\brak{\RR\cdot\intvol{\BB}{\rho\phibr}} 
  \label{eq:total_momentum__angular} .
\end{align}
\end{subequations}
These expressions, which are equivalent to eq.~\eqref{eq:gen_mom_JJ_euc}, clarify that we still have not achieved a `minimal coupling' of the internal and Euclidean motions inasmuch as \( \mom \) and \( \angmom \) still depend on \( \phibr \). Fortunately though, they depend on \( \phibr \) only through \( \intvol{\BB}{\rho\phibr} \) which, courtesy of constraint~\eqref{eq:constraint_explicit_tmp__P}, is constant in time and therefore only represents three degrees of freedom. We therefore declare that
\begin{align}
\label{eq:constraint_new_com}
  \intvol{\BB}{\rho\phibr} = 0 ,
\end{align} 
and will now show that this is allowed by our freedom to choose initial conditions on \( \Phib \) and \( \RR \).

Given definite initial conditions on the spatial motion \( \phib(\xx,0) = \phib_{0} \) (the initial condition on the velocity is irrelevant at present) we trivially have
\begin{align}
  \label{eq:init_cond_explicit_tmp__phib}
  \phib_{0} 
  &= 
  \Phib_{0} + \RR_{0}\cdot\phibr(\xx,0),
\end{align}
where \( \Phib_{0} \) and \( \RR_{0} \) are (so far) arbitrary initial conditions on \( \Phib \) and \( \RR \). Integrating over \( \BB \) gives
\begin{align}
  \intvol{\BB}{\rho\phib_{0}}
  =
  m\Phib_{0}
  +
  \RR_{0}\cdot\intvol{\BB}{\rho\phibr(\xx,0)} .
\end{align}
It follows that a necessary and sufficient condition for eq.~\eqref{eq:constraint_new_com} to hold is
\begin{align}
\label{eq:com_ajwoiehjkepio2e}
  \Phib_{0} = \frac{1}{m}\intvol{\BB}{\rho\phib(\xx,0)} \equiv \CoM ,
\end{align}
\ie that \( \Phib_{0} \) gives the initial position of the body's CoM; \( \RR_{0} \) remains arbitrary. As long as we use these initial conditions, then we may replace constraint~\eqref{eq:constraint_explicit_tmp__P} with the stronger constraint~\eqref{eq:constraint_new_com} henceforward. 

This new constraint means that \( \Phib \) is the CoM of the body at all times and carries all the body's linear momentum: eq.~\eqref{eq:total_momentum__linear} becomes
\begin{align} 
  \mom = m\VV
  \label{eq:P_Phib_gauge},
\end{align} 
which we may integrate trivially to give
\begin{align}
  \Phib = \CoM + \frac{1}{m}\mom t .
\end{align}
The total angular momentum also 
factors neatly into the \textit{orbital} angular momentum of the CoM's linear motion,  
and \textit{spin} angular momentum associated with rotation about the CoM:
\begin{align}
\label{eq:ang_mom_gauge}
  \angmom
  &=
  \mom\wedge\CoM
  +
  \II\cdot\Omb .
\end{align}
With the moment of inertia and angular velocity as defined, the \textit{form} of the spin angular momentum is even the same as for a rigid body, but with
  an `effective'
moment of inertia that accounts for the body's deformation.

\subsection{New variational principle}
\label{sec:var_prin_solid_decomp}

\subsubsection{Decomposed action}

We now substitute eq.~\eqref{eq:decomposition} into the action in order to obtain a variational principle that depends on both the Euclidean and internal motions. From now on we will use `internal' expressions almost exclusively. Using eqs.~(\ref{eq:ur_reconstruction_equation_internal}, \ref{eq:velocity_diff__2}) to write
\begin{align}
  \norm{\vv}^{2}
  &=
  \norm{\vvr+\eucr\cdot\phibr}^{2}
  =
  \norm{\vvr}^{2}
  +
  2\ip{\vvr}{\eucr\cdot\phibr}
  +
  \ip{\eucr\cdot\phibr}{\eucr\cdot\phibr} ,
\end{align}
we readily find from 
eqs.~(%
  \ref{eq:elastic_body_basic_action__grav_take_3}, 
  \ref{eq:effective_mass_defn__internal},
  \ref{eq:pi_internal_defn})
that the action~\eqref{eq:elastic_body_basic_action__grav_take_3} decomposes as
\begin{align}
\label{eq:action_decomposed_initial}
  \action 
  &= 
  \inttime{\III}{\frac{1}{2}\fip{\JJr\cdot\eucr}{\eucr}}
  +
  \inttime{\III}{\fip{\pibr}{\eucr}}
  +
  \inttime{\III}{
    \intvol{\BB}{\braksq{
   		 \frac{1}{2}\rho\norm{\vvr}^{2}
   		-W\brak{\xx,t,\FFr}
		-\rho\,\zeta
    }}
  }
  \nonumber\\
  &\qquad
  -
  \frac{1}{8 \pi G}
  \inttime{\III}{
    \intvol{\RThree}{
      \ip{\nabla\zeta}{\aar\cdot\nabla\zeta} 
    }
  } ,
\end{align}
with \( \FFr \) the deformation gradient of \( \phibr \) and 
\begin{align}
  \aar \equiv \brak{\det{\FFr}}\FFr^{-1}\FFr^{-T} .
\end{align}
The original constraint \( \pibr = 0 \) knocks out the second term, while the choice of \( \Phib_{0} \) eliminates the cross-terms between \( \vvr \) and \( \Ombr \) in the kinetic energy:
\begin{align}
\label{eq:JJr_gauge}
  \frac{1}{2}\fip{\JJr\cdot\eucr}{\eucr}
  &=
  \frac{1}{2}m\norm{\RR^{T}\cdot\VV}^{2}
  +
  \intvol{\BB}{\rho\ip{\RR^{T}\cdot\VV}{\Ombr\cdot\phibr}}
  +
  \frac{1}{2}\ip{\Ombr}{\IIr\cdot\Ombr}
  \nonumber\\
  &=
  \frac{1}{2}m\norm{\VV}^{2}
  +
  \frac{1}{2}\ip{\Ombr}{\IIr\cdot\Ombr} .
\end{align}
Thus we may write the action as 
\begin{align}
  \action 
  &=
  \inttime{\III}{\frac{1}{2}m\norm{\VV}^{2}}
  +
  \inttime{\III}{\frac{1}{2}\ip{\Ombr}{\IIr\cdot\Ombr}}
  \nonumber\\
  &\qquad\qquad
  +
  \inttime{\III}{
    \intvol{\BB}{\braksq{
   		 \frac{1}{2}\rho\norm{\vvr}^{2}
   		-W\brak{\xx,t,\FFr}
		-\rho\,\zeta
    }}
  }
  -
  \frac{1}{8 \pi G}
  \inttime{\III}{
    \intvol{\RThree}{
      \ip{\nabla\zeta}{\aar\cdot\nabla\zeta} 
    }
  } ,
\end{align}
subject to the constraints
\begin{subequations}
\label{eq:decomposition_constraint_full}
\begin{align}
  &\intvol{\BB}{\rho\phibr} = 0
  \\
  &\intvol{\BB}{\rho\vvr\wedge\phibr} = 0
\end{align}
\end{subequations}
and with the motion given by eq.~\eqref{eq:decomp_explicit__phi}.
Intuitively, we can see eqs.~\eqref{eq:decomp_explicit__phi}~and~\eqref{eq:decomposition_constraint_full} as describing elastic motion that takes place in a rotating frame whose orientation is given by \( \RR \) relative to an origin at \( \Phib \), with the elastic motion within the rotating frame carrying neither linear nor angular momentum and with its CoM always 
  at the origin.

We could of course have chosen the constraints differently,
but the choice we have made seems to be the most convenient.
Indeed, the action has factored into a `fully internal' part (the third and fourth terms) cosmetically identical to eq.~\eqref{eq:elastic_body_basic_action__grav_take_3}'s action, but now augmented by two terms which feature the Euclidean motion. Those first two terms are  identical in form to a rigid body's action \citep[\eg][]{holm2009geometric}, but with the one subtle difference that \( \IIr \) here is not constant due to its dependence on \( \phibr \).
Most importantly, the constraints ensure that \( \phibr \) can have no secular behaviour.

\subsubsection{The reconstruction equation}
\label{sec:reduction}

The decomposed action has no explicit dependence on the Euclidean motion. \( \Euc \) only enters implicitly via \( \eucr \equiv \Euc^{-1}\dot{\Euc} \). The variational principle will provide differential equations for \( \eucr \), \( \phibr \) and \( \zeta \), but not for \( \Euc \). In order to write down the full, spatial motion \( \phib = \Euc\cdot\phibr \), we need to be able to relate \( \Euc \) to \( \eucr \), \( \phibr \) and \( \zeta \). That relationship is the \textit{reconstruction equation}
\begin{subequations}
\label{eq:recons_full}
\begin{align}
  &\dot{\Euc} = \Euc\eucr
  \label{eq:recons_full__eq} \\
  &\Euc(0)    = \Euc_{0}
  \label{eq:recons_full__init} ,
\end{align}
\end{subequations}
an ODE on \( \EThree \) obtained by rearranging eq.~\eqref{eq:ur_reconstruction_equation_internal}. To couch eq.~\eqref{eq:recons_full} in terms of \( \Phib \), \( \RR \) etc., we use the identities
\begin{subequations}
\begin{align}
  \Euc\eucr\cdot\aa
  &=
  \RR\cdot\brak{\RR^{T}\cdot\VV
                +
                \Ombr\cdot\aa 
          }
  \\
  \dot{\Euc}\cdot\aa 
  &=
  \VV + \dot{\RR}\cdot\aa
\end{align}
\end{subequations}
for arbitrary \( \aa \in \RThree \). It follows that eq.~\eqref{eq:recons_full} implies the trivial ODE \( \dot{\Phib} = \VV \), as well as
\begin{subequations}
\label{eq:reconstruction_equation}
\begin{align}
  &\dot{\RR} = \RR\Ombr
  \\
  &\RR(0) = \RR_{0} .
\end{align}
\end{subequations}
Henceforward we refer to eq.~\eqref{eq:reconstruction_equation} without ambiguity as ``the reconstruction equation" -- and we emphasise that \( \RR_{0} \) may still be chosen according to convenience.

The reconstruction equation is a kinematic identity that may be stated without reference to the internal dynamics. And as alluded to just above, we see concretely that \( \RR \) only enters via an ODE once the 
  rest of the equations
have been solved and \( \Ombr \) is known fully. We do not have to integrate the reconstruction equation simultaneously with the others. Nevertheless, eq.~\eqref{eq:reconstruction_equation} acts as a further constraint on the action that must be accounted for when deriving equations of motion.

\subsubsection{Constraining the action}

We will impose the reconstruction equation \textit{strongly}, that is without the use of Lagrange multipliers. On the other hand we choose to impose the Euclidean constraints weakly, introducing the Lagrange multipliers
\begin{subequations}
\begin{align}
  \alphab &: \III \rightarrow \RThree  \\
  \betab  &: \III \rightarrow \sothree .
\end{align}
\end{subequations}
The action now takes the final form
\begin{align}
\label{eq:action_variably_rotating}
  \action
  &= 
  \inttime{\III}{\frac{1}{2}m\|\VV\|^{2}}
  +
  \inttime{\III}{\frac{1}{2}\ip{\Ombr}{\IIr\cdot\Ombr}}
  \nonumber\\
  &\qquad\qquad
  +
  \inttime{\III}{
    \intvol{\BB}{\braksq{
   		 \frac{1}{2}\rho\norm{\vvr}^{2}
   		+\rho\ip{\alphab}{\phibr}
   		+\rho\ip{\vvr}{\betab\cdot\phibr}
   		-W\brak{\xx,t,\FFr}
		-\rho\,\zeta
    }}
  }
  \nonumber\\
  &\qquad\qquad
  -
  \frac{1}{8 \pi G}
  \inttime{\III}{
    \intvol{\RThree}{
      \ip{\nabla\zeta}{\aar\cdot\nabla\zeta} 
    }
  } .
\end{align}
The action is to be viewed as a functional of \( \VV \), \( \Ombr \), \( \phibr \) and \( \zeta \), as well as \( \alphab \) and \( \betab \).

We have chosen to impose the constraints in this mixed weak--strong way for the following reasons. Regarding \( \Ombr \), it is easy to impose its constraint~\eqref{eq:reconstruction_equation} directly upon the variation \( \delta\Ombr \) at the point of applying Hamilton's principle, so there is no need for a Lagrange multiplier (see below, Section~\ref{sec:variable_rotation_action_variation}). The analogous procedure does not work (not to our knowledge, at least) for the angular-momentum constraint, so that must be applied through a Lagrange multiplier. If only for aesthetic reasons we impose the translational constraint the same way. 

Finally, for later reference we define the `internal elastic action'
\begin{align}
\label{eq:internal_elastic_action_defn}
  \actionSmall
  &=
  \inttime{\III}{
    \intvol{\BB}{\braksq{
   		 \frac{1}{2}\rho\norm{\vvr}^{2}
   		+\rho\ip{\alphab}{\phibr}
   		+\rho\ip{\vvr}{\betab\cdot\phibr}
   		-W\brak{\xx,t,\FFr}
		-\rho\,\zeta
    }}
  }
  \nonumber\\
  &\qquad\qquad
  -
  \frac{1}{8 \pi G}
  \inttime{\III}{
    \intvol{\RThree}{
      \ip{\nabla\zeta}{\aar\cdot\nabla\zeta} 
    }
  } ,
\end{align}
in terms of which \( \action \) takes the concise form
\begin{align}
  \action 
  = 
  \inttime{\III}{\frac{1}{2}m\|\VV\|^{2}}
  +
  \inttime{\III}{\frac{1}{2}\ip{\Ombr}{\IIr\cdot\Ombr}}
  +
  \actionSmall .
\end{align}

\subsection{Equations of motion}
\label{sec:single_body_eoms}

We are now ready to derive the differential equations governing the Euclidean and internal motions. As a final piece of pragmatism we will drop all the overlines on internal variables henceforward, reinstating them only on the rare occasions that they are needed to avoid ambiguity.

\subsubsection{Variation of the action}
\label{sec:variable_rotation_action_variation}

In order to obtain the EoMs we vary with respect to all variables 
and demand that the first variation of the action vanish. Starting with the Lagrange multipliers, variation in \( \alphab \) and \( \betab \) trivially returns the Euclidean constraints,
\begin{subequations}
\label{eq:variable_rotation_eom_euclid}
\begin{align}
  &\intvol{\BB}{\rho\phib} = 0 \\
  &\intvol{\BB}{\rho\vv\wedge\phib} = 0 .
\end{align}
\end{subequations}
Varying with respect to \( \Phib \) is almost as easy:
we find that
\begin{align}
  \delta\action
  =
  \inttime{\III}{m\ip{\VV}{\partial_{t}\delta\Phib}},
\end{align}
from which we obtain the ODE
\begin{align}
\label{eq:variable_rotation_eom_newton}
  m\dot{\VV} = 0.
\end{align}
As expected for a system with no external forcing, the CoM moves at constant velocity.

Variation with respect to \( \Omb \) is more involved because the variation \( \delta\Omb \in \sothree \) cannot be chosen freely due to the constraint \( \Omb = \RR^{T}\dot{\RR} \). Instead we must consider arbitrary variations in \( \RR \).
We may write a small change in \( \RR \) as
\begin{align}
  \RR \rightarrow \RR\ee^{\delta\thetab}
\end{align}
for arbitrary small \( \delta\thetab \in \soLie(3) \); this way the group structure is respected and the right hand side is guaranteed to be an element of \( \SOThree \). The exponential is then Taylor-expanded to give
\begin{align}
\label{eq:rigid_body_R_variation}
  \RR &\rightarrow \RR + \RR\delta\thetab .
\end{align}
For the variation of \( \Omb \) we therefore have
\begin{align}
  \delta\Omb 
  &\equiv \delta\RR^{T}\dot{\RR} + \RR^{T}\delta\dot{\RR}
  \nonumber\\
  &= \brak{\RR\delta\thetab}^{T}\dot{\RR} + \RR^{T}\partial_{t}\brak{\RR\delta\thetab}
  \nonumber\\
  &= \brak{\partial_{t} + \smallAd{\Omb}}\cdot\delta\thetab ,
\end{align}
where the \textit{little-adjoint} \(\smallAd{}\) is defined in eq.~\eqref{eq:rigid_body:small_ad_defn}.
We can schematically write any term in the action involving \( \Omb \) as \( \inttime{\III}{f(\Omb)} \) for some 
scalar-valued function \( f : \soThree \mapsto \III \); recalling that all fields are varied subject to fixed endpoint conditions, it then follows that any term in the action involving \( \Omb \) satisfies
\begin{align}
  \inttime{\III}{f\brak{\Omb+\delta\Omb}}
  &= 
  \inttime{\III}{\braksq{
    f\brak{\Omb} 
    + 
    \ip{D f}{\brak{\partial_{t} + \smallAd{\Omb}}\cdot\delta\thetab}
  }}
  \nonumber\\
  &=
  \inttime{\III}{\braksq{
    f\brak{\Omb} 
    - 
    \ip{\brak{\partial_{t} + \smallAd{\Omb}}\cdot D f}{\delta\thetab}
  }} 
  ,  
\end{align}
so that the corresponding variation is
\begin{align}
  \delta\inttime{\III}{f\brak{\Omb}}
  &=
  -\inttime{\III}{
  \ip{\brak{\partial_{t} + \smallAd{\Omb}}\cdot D f}{\delta\thetab}
  } .
\end{align}
For the present case \( f(\Omb) = \frac{1}{2}\ip{\Omb}{\II\cdot\Omb} \), so this implies
\begin{align}
  \delta\action
  =
  -\inttime{\III}{\ip{\delta\thetab}{\brak{\partial_{t} + \smallAd{\Omb}}\II\cdot\Omb}} .
\end{align} 
We have derived the equation
\begin{align}
\label{eq:variable_rotation_eom_euler}
  \brak{\partial_{t} + \smallAd{\Omb}}\II\cdot\Omb = 0 ,
\end{align}
which is the celebrated \textit{Euler equation} of rigid body dynamics, but with a body MoI that depends on \( \phib \) and therefore on time.

Next, variation with respect to the referential potential is accomplished readily:
\begin{align}
  -\inttime{\III}{\intvol{\BB}{\rho\,\delta\zeta}}
  -\frac{1}{4\pi G}
   \inttime{\III}{\intvol{\RThree}{
     \ip{\aa\cdot\nabla\zeta}{\nabla\delta\zeta}
   }}
   =0 .
\end{align}
Noting that this expression contains no time-derivatives of the variation \( \delta\zeta \), it is equivalent to
\begin{align}
\label{eq:variable_rotation_eom_poisson}
   \intvol{\BB}{\rho\,\delta\zeta}
  +\frac{1}{4\pi G}
   \intvol{\RThree}{
    \ip{\aa\cdot\nabla\zeta}{\delta\zeta}
   }
  =0 ,
\end{align}
where \( \delta\zeta \) is a time-independent test-function. This is of course the referential Poisson equation in weak-form.

Now for variation in \( \phib \). 
From definition~\eqref{eq:moi_defn_int}, the MoI's variation satisfies
\begin{align}
  \ip{\Omb}{\delta\II\cdot\Omb}
  &=
  \ip{
    \Omb
  }{
    \intvol{\BB}{\rho\braksq{
      \brak{\Omb\cdot\phib}\wedge\delta\phib
      +
      \brak{\Omb\cdot\delta\phib}\wedge\phib
    }}
  }
  =
  \intvol{\BB}{\rho\ip{\delta\phib}{-2\Omb^{2}\cdot\phib}} ,
\end{align}
so
\begin{align}
  &
  \inttime{\III}{\intvol{\BB}{\braksq{
    \rho\ip{\vv+\betab\cdot\phib}{\delta\vv}
    -
    \rho\ip{\alphab+\betabdot\cdot\phib+\Omb^{2}\cdot\phib}{\delta\phib}
    -
    \ip{\TT}{\delta\FF}
  }}}
  \nonumber\\
  &\qquad
  -
  \frac{1}{8 \pi G}
  \braksq{
    \inttime{\III}{\intvol{\RThree}{
      \ip{\nabla\zeta}{\delta\aa\cdot\nabla\zeta}
    }}
  }
  =
  0 ,
\end{align}
with 
\begin{align}
  \delta\vv &= \partial_{t}\delta\phib \\
  \delta\FF &= D_{\xx}\delta\phib
\end{align}
and where \( \TT \) is the (internal) \textit{first Piola--Kirchoff stress} 
\begin{align}
\label{eq:fpk_defn}
  \TT \equiv D_{\FF}W .
\end{align}
The analogous \textit{first Piola--Kirchoff gravitational stress} is
\begin{align}
\label{eq:grav_stress_defn}
  \NNN \equiv D_{\FF}\brak{\frac{1}{8\pi G}\ip{\nabla\zeta}{\aa\cdot\nabla\zeta}} .
\end{align}
If we define the \textit{referential gravitational acceleration}
\begin{align}
  \gam = -\FF^{-T}\cdot\nabla\zeta ,
\end{align}
then \( \NNN \) is given explicitly by
\begin{align}
\label{eq:grav_stress_defn__explicit}
  \NNN
  &\equiv
  \frac{1}{8 \pi G}
  J\brak{\norm{\gam}^{2}\id- 2 \gam\otimes\gam}\FF^{-T} , 
\end{align} 
with \(\id\) the identity tensor. 
It is convenient to split the integral over \( \RThree \) into respective integrals over \( \BB \) and \( \RThree/\BB \):
\begin{align}
  \inttime{\III}{\intvol{\BB}{\braksq{
    \rho\ip{\vv+\betab\cdot\phib}{\delta\vv}
    -
    \rho\ip{\alphab+\betabdot\cdot\phib+\Omb^{2}\cdot\phib}{\delta\phib}
    -
    \ip{\TT+\NNN}{\delta\FF}
  }}}
  -
  \inttime{\III}{\intvol{\RThree/\BB}{
    \ip{\NNN}{\delta\FF}
  }}
  =
  0 .
\end{align} 
The exterior term may be integrated by parts to give
\begin{align}
  \inttime{\III}{\intvol{\RThree/\BB}{
    \ip{\NNN}{\delta\FF}
  }}
  =
  -
  \inttime{\III}{\intsurf{\partial\BB}{
    \ip{\NNN\cdot\nhat}{\delta\phib}
  }}
  -
  \inttime{\III}{\intvol{\RThree/\BB}{
    \ip{\Div\NNN}{\delta\phib}
  }} .
\end{align}
By direct calculation one may show that \(\Div\NNN = \rho\gam\) (\cf D\&T, Section~2.8.3), so the second term on the RHS vanishes. With this, the last step is to note that \( \delta\phib \) vanishes at the endpoints and time-integrate the \( \delta\vv \) term by parts.
This results in the weak-form momentum equation
\begin{align}
\label{eq:variable_rotation_eom_momentum_initial}
  &
  \intvol{\BB}{
    \rho
    \ip{
       \dot{\vv}
      +\Omb^{2}\cdot\phib
      +\partial_{t}\brak{\betab\cdot\phib}
      +\betabdot\cdot\phib
      +\alphab
    }{
      \delta\phib
    }
  }
  +
  \intvol{\BB}{
    \ip{\TT+\NNN}{\delta\FF}
  }
  +
  \intsurf{\partial\BB}{
    \ip{\NNN\cdot\nhat}{\delta\phib}
  }
  =0  ,
\end{align}
where \( \delta\phib \) now denotes a time-independent test-function.

\subsubsection{Identification of the Lagrange multipliers}
\label{sec:lag_mult_ident_solid}

To complete the derivation we eliminate the Lagrange multipliers \( \alphab \) and \( \betab \). 
To deal with \( \alphab \), let \( \delta\phib = \aa \) within eq.~\eqref{eq:variable_rotation_eom_momentum_initial} for arbitrary \( \aa\in\RThree \). Due to the translational constraint on \( \phib \) we then have
\begin{align}
  \intvol{\BB}{\rho\ip{\alphab}{\aa}}
  =
  0
\end{align}
whence \( \alphab \) vanishes.

For \( \betab \) take
\( \delta\phib = \thetab\cdot\phib \) so that
\( \delta\FF = \thetab\FF \)
for arbitrary \( \thetab\in\soThree \). The FPK stress term vanishes because
\begin{align}
  \intvol{\BB}{\ip{\TT}{\thetab\FF}}
  =
  \intvol{\BB}{\ip{\TT\FF^{T}}{\thetab}}
\end{align}
and \( \TT \) is such that \( \TT\FF^{T} \) is symmetric; an analogous result holds for the gravitational stress. The kinetic term vanishes because of the rotational constraint (and \( \thetab \)'s antisymmetry):
\begin{align}
  \intvol{\BB}{\ip{\rho\dot{\vv}}{\thetab\cdot\phib}}
  =
  \dbyd{t}\intvol{\BB}{\ip{\rho\vv}{\thetab\cdot\phib}}
  =
  \ip{\thetab}{\dbyd{t}\intvol{\BB}{\rho\vv\wedge\phib}} .
\end{align}
Meanwhile the centrifugal term satisfies
\begin{align}
  \intvol{\BB}{\rho\ip{\Omb^{2}\cdot\phib}{\thetab\cdot\phib}}
  =
  \ip{\smallAd{\Omb}\II\cdot\Omb}{\thetab},
\end{align}
and, after a little algebra, one may show that
\begin{align}
  \intvol{\BB}{\rho\ip{\partial_{t}\brak{\betab\cdot\phib}+\betab\cdot\vv}{\thetab\cdot\phib}}
  =
  \ip{\partial_{t}\brak{\II\cdot\betab}}{\thetab} .
\end{align}
We conclude that the Lagrange multiplier satisfies
\begin{align}
  \partial_{t}\brak{\II\cdot\betab}+\smallAd{\Omb}\II\cdot\Omb = 0 .
\end{align}
Given eq.~\eqref{eq:variable_rotation_eom_euler}, one may obtain a solution to the present equation by setting
\begin{align}
  \betab = \Omb .
\end{align}

\subsubsection{Summary}

We can summarise the results of the preceding derivation as follows:
\begin{subequations}
\label{eq:eoms_variable_rotation_weak_form_final}
\begin{align}
  &
  \dot{\RR} = \RR\Omb 
  \label{eq:eoms_variable_rotation_weak_form_final__recons}\\
  &
  m\dot{\VV} = 0
  \label{eq:eoms_variable_rotation_weak_form_final__com}\\
  &
  \brak{\partial_{t} + \smallAd{\Omb}}\II\cdot\Omb = 0
  \\
  &
  \intvol{\BB}{
    \rho
    \ip{
       \dot{\vv}
      +\Omb^{2}\cdot\phib
      +2\Omb\cdot\vv
      +\dot{\Omb}\cdot\phib
    }{
      \ww
    }
  }
  +
  \intvol{\BB}{
    \ip{\TT+\NNN}{D\ww}
  }
  +
  \intsurf{\partial\BB}{
    \ip{\NNN\cdot\nhat}{\ww}
  }
  = 0
  \label{eq:eoms_variable_rotation_weak_form_final__mom}\\
  &
  \intvol{\BB}{\rho\potweak}
  +
  \frac{1}{4\pi G}
  \intvol{\RThree}{
    \ip{\aa\cdot\nabla\zeta}{\nabla\potweak}
  }
  =0 
  \label{eq:eoms_variable_rotation_weak_form_final__poisson}\\
  &
  \intvol{\BB}{\rho\phib} = 0 
  \\
  &
  \intvol{\BB}{\rho\vv\wedge\phib} = 0 .
\end{align}
\end{subequations}
We have renamed \( \delta\phib \) to \( \ww \) and \( \delta\zeta \) to \( \potweak \) in order to help clarify that we now regard them as arbitrary test-functions. The first three equations describe the evolution of the Tisserand frame. We find that the CoM obeys Newton's Second Law, while the angular velocity obeys an Euler equation with a time-dependent MoI. The reconstruction equation~\eqref{eq:eoms_variable_rotation_weak_form_final__recons} is a kinematic identity with no influence on the solutions for the other variables.
The internal motion itself obeys a momentum equation reminiscent of the standard equations of global seismology, but with some important differences. Firstly, there is no requirement that the angular velocity in eq.~\eqref{eq:eoms_variable_rotation_weak_form_final__mom} be either small or constant: this equation describes finite elastodynamics in a variably rotating frame. It couples to the equations describing the frame's evolution through \(\Omb\) and \(\II\). It also couples to the Poisson equation~\eqref{eq:eoms_variable_rotation_weak_form_final__poisson} through \(\phib\) and \(\zeta\). The Euclidean constraints are imposed on the motion strongly and, thanks to \( \alphab \) and \( \betab \), on the test-functions implicitly. 
By solving for the Lagrange multipliers we explicitly recovered the Coriolis and Euler forces.

It is worth commenting that we have left eqs.~\eqref{eq:eoms_variable_rotation_weak_form_final__mom} and \eqref{eq:eoms_variable_rotation_weak_form_final__poisson} in weak-form. If the corresponding strong-form equations are ever needed they could be derived readily, but weak-forms are generally more useful for numerical work and are therefore our focus here and elsewhere in the paper.

\section{Extension to fluid--solid planets}
\label{sec:fluid_solid}

If we wish to model Earth-like planets, then we must adapt the preceding approach to account for (visco)elastic bodies composed of alternating fluid and solid layers. That is the subject of Sections \ref{sec:fluid_solid_kinematics}--\ref{sec:fluid_solid_body_eoms_exact}. Section~\ref{sec:vorticity_maybe} discusses a symmetry group associated with fluid regions along with the associated conservation law.

\subsection{The form of the fluid strain-energy function}
\label{sec:fluid_solid_strain_energy}

Within an \textit{elastic fluid}, the strain-energy function takes the form
\citep[\eg][]{Noll_1974}
\begin{align}
\label{eq:fs_strain_energy}
    W(\xx,\FF) = V(\xx,J)
\end{align}
for some function \( V \) and where we recall that \( J = \det{\FF} \). It follows that the FPK stress is
\begin{align}
    \TT 
    &=
    - p J \FF^{-T} ,
\end{align}
where we have defined the \textit{pressure}
\begin{align}
\label{eq:fs_p_defn}
    p = -D_{J}V
\end{align}
and made use of \textit{Jacobi's formula}
\citep[\eg][]{Holzapfel_2000}
for the derivative of a determinant.

\subsection{Tangential slip}
\label{sec:fluid_solid_kinematics}

\begin{figure}
  \centering
  \includegraphics[width=0.8\textwidth]{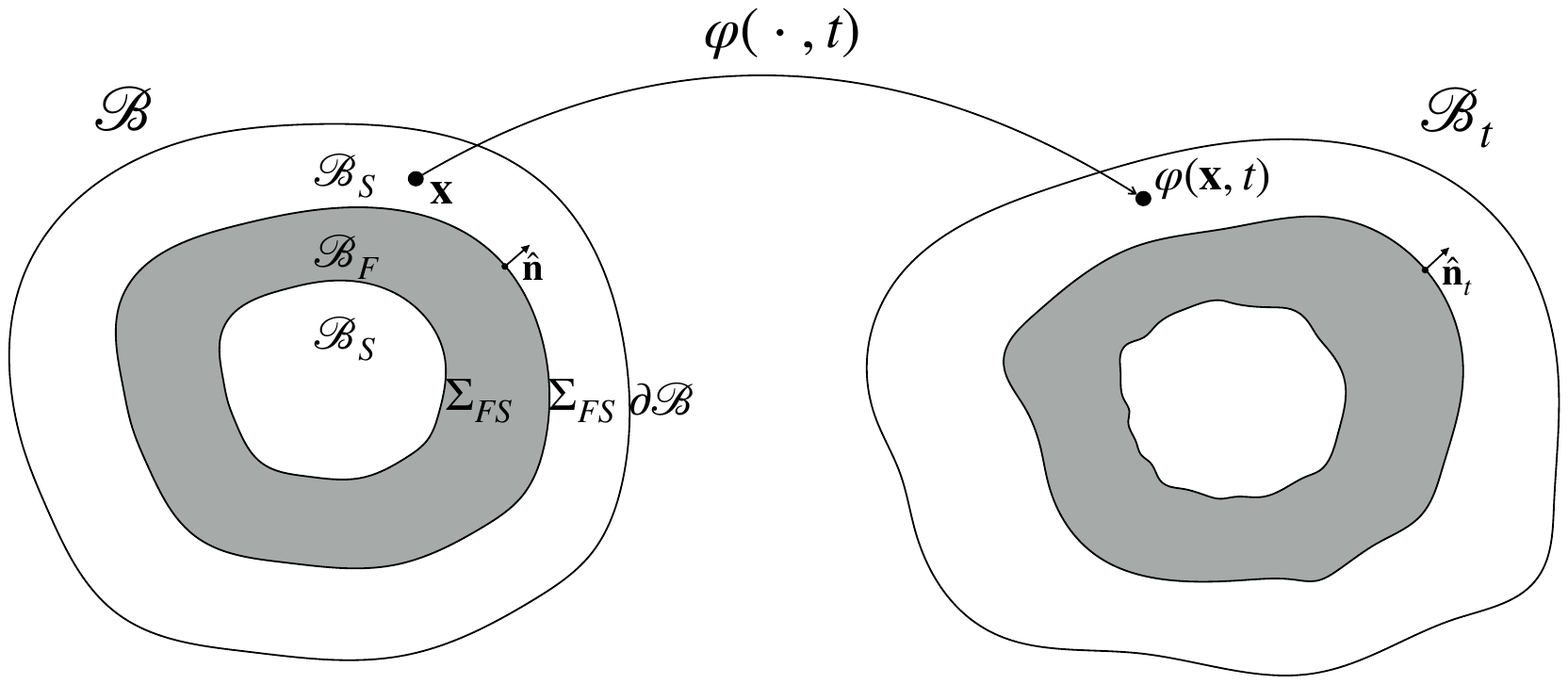}
  \caption{A three-layer fluid-solid planet (after D\&T, Fig.~3.1). An arbitrary point \( \xx \) in the reference body \( \BB \) is mapped to \( \phib(\xx,t) \) in the instantaneous configuration \( \BB_{t} \). \( \BB_{S} \) and \( \BB_{F} \) respectively denote the solid and fluid regions of the reference body. \( \Sigma_{FS} \) refers to the union of all internal fluid--solid boundaries, while \( \partial\BB \) is the free-surface. A unit vector \( \nhat \) on a boundary is mapped by \( \phib \) to \( \nhat_{t} \propto J \FF^{-T}\cdot\nhat \). In this illustration the uppermost layer is solid, but this Section's theoretical developments do not require that.}
  \label{fig:N_body}
\end{figure}

Consider a body composed of \( N \) distinct layers, alternating between fluid and solid; Fig.~\ref{fig:N_body} shows a special case consisting of three layers.
It is not relevant at present which of the layers are fluid, only that the layers possess mutual fluid--solid boundaries where the fluid is able to slide freely past the solid. Such \textit{tangential slip} is the crucial kinematical difference between solid and fluid--solid bodies. Otherwise all of 
our earlier
kinematical considerations carry over to a layered fluid--solid planet.

\subsubsection{Illustrative special case: two layers}

We assume that the reference body \( \BB \) contains a single referential fluid--solid boundary \( \SigFS \). 
Consider two particles on either side of \( \SigFS \) that are instantaneously adjacent at some time. During the subsequent motion of the body these particles can separate from one another as the fluid slides tangentially past the moving solid interface. The motion of such a body can therefore be discontinuous across its fluid--solid boundaries, but as long as the layers neither separate nor interpenetrate, the motion is subject to the constraint that
\begin{align}
  \phib_{-}\brak{\SigFS,t} = \phib_{+}\brak{\SigFS,t} ,
\end{align}
where \( \phib_{-} \) (resp. \( \phib_{+} \)) gives the motion evaluated immediately below (resp. above) \( \SigFS \). This relation does not require the point-wise equality \( \phib_{-}(\xx,t) = \phib_{+}(\xx,t) \), but only the weaker condition that for each \( \xx\in\SigFS \) there is at time \( t \) a unique point \( \xx'\in\SigFS \) such that \( \phib_{-}(\xx,t) = \phib_{+}(\xx',t) \). Put differently,
\begin{align}
\label{eq:tangential_slip_ur_ur_tmp}
  \phib_{-}\circ\chib = \phib_{+}
\end{align}
for some \( \chib \in \DIFF{\SigFS} \). Due to this constraint, the configuration space for the body is no longer
the diffeomorphism group, \(\DIFF{}\), but must be  extended to allow for suitable  piecewise-diffeomorphisms.


\subsubsection{Extension to \( N \) layers}

For an \( N \)-layer planet the reference body \( \BB \) is decomposed in the form
\begin{align}
  \BB = \BB_{1}\cup\BB_{2}\cup\dots\cup\BB_{N}
\end{align}
with referential fluid--solid boundaries
\begin{align}
  \Sigma_{i} = \BB_{i}\cap\BB_{i+1} \quad i<N .
\end{align}
Following D\&T, we now use the symbol \( \SigFS \) to denote the \textit{union} of those boundaries:
\begin{align}
  \SigFS = \Sigma_{1}\cup\dots\cup\Sigma_{N-1} .
\end{align}
The same kinematic considerations as above hold for each of the fluid--solid boundaries individually, so the motion's restriction to \( \SigFS \) must satisfy
\begin{align}
\label{eq:tangential_slip_ur_tmp_2}
  \phib_{-}\circ\chib = \phib_{+} ,
\end{align}
where once again \( \chib \in \DIFF{\SigFS} \) and we use a subscript plus and minus to denote evaluation of \( \phib \) just above and below \( \SigFS \). This notation allows us to deal with any number of fluid--solid boundaries without any cosmetic difference from the case of a single such boundary. In particular, it avoids the need to explicitly decompose \( \BB \) into multiple sub-bodies. 
Nevertheless, in Sections~\ref{sec:vorticity_maybe} and \ref{sec:equilibrium_configs} it will be helpful to consider \( \BB \)'s fluid and solid regions separately. To that end we import from D\&T the notation \( \BB_{F} \) (resp. \( \BB_{S} \)) to denote the union of all of \( \BB \)'s fluid (resp. solid) regions, with \( \BB = \BB_{F}\cup\BB_{S} \).

\subsection{Variational principle (or, `Everything else is basically the same')}
\label{sec:f_s_var_prin}

There is no potential energy associated specifically with such tangential slip at a fluid--solid boundary. Although the tangential slip modifies the problem's configuration space, the action of a fluid--solid body is therefore unchanged from eq.~\eqref{eq:elastic_body_basic_action__grav_take_3}.
Indeed, the tangential slip constraint has no impact on the derivations of Sections
\ref{sec:symmetries}, 
\ref{sec:solid_noether} and
\ref{sec:decomposition_solid}.
The action is of course subject to constraint~\eqref{eq:tangential_slip_ur_tmp_2}, but it still exhibits the same symmetries under left- and right-actions of \( \DIFF{} \) and \( \EThree \); the body's total momentum and energy are conserved; and the decomposition
\begin{subequations}
\label{eq:decomp_fs}
\begin{align}
  &\phib = \Phib + \RR\cdot\phibr 
  \\
  &\intvol{\BB}{\rho\phibr} = 0
  \\
  &\intvol{\BB}{\rho\vvr\wedge\phibr} = 0
\end{align}
\end{subequations}
may clearly still be made without violating eq.~\eqref{eq:tangential_slip_ur_tmp_2}. In fact, the tangential slip constraint on \( \phib \) takes the same form when applied to \( \phibr \):
\begin{align}
\label{eq:tangential_slip_ur}
  \phibr_{-}\circ\chib = \phibr_{+} ,
\end{align}
for some \( \chib \in \DIFF{\SigFS} \). We now revert to notating the internal motion by \( \phib \) and the body angular velocity by \( \Omb \).

It is crucial to recognise that both the motion and the test-functions are constrained, and moreover that the test-functions satisfy a \textit{different constraint from the motion}. Not only do we have
\begin{align}
\label{eq:tangential_slip_constraint_motion}
  \phib_{-}\circ\chib = \phib_{+}
\end{align}
on \( \SigFS \), but also, taking this equation's first-variation,
\begin{align}
  \delta\phib_{-}\circ\chib
  +
  \brak{\FF_{-}\circ\chib}\cdot\delta\chib
  =
  \delta\phib_{+} .
\end{align}
Since \( \chib \in \DIFF{\SigFS} \), \( \delta\chib \) is tangent to the fluid--solid boundary at \(\chib(\xx\)) (see A18,~eq.~55), so this is equivalent to
\begin{align}
\label{eq:tangential_slip_constraint_initial}
  \ip{
    \brak{\FF_{-}\circ\chib}^{-1}
    \cdot
    \brak{\delta\phib_{+}-\delta\phib_{-}\circ\chib}
  }{
    \nhat_{-}\circ\chib
  }
  =0 ,
\end{align}
with \( \nhat_{-}(\xx) \) the outward-pointing normal to \( \SigFS \) at \( \xx \) (that is, the outward-pointing normal with respect to the sub-body just \textit{beneath} the boundary).
This translates straight into the constraint satisfied by the test-functions:
\begin{align}
\label{eq:tangential_slip_constraint_test_functions}
  \ip{
    \brak{\FF_{-}\circ\chib}^{-1}
    \cdot
    \brak{\ww_{+}-\ww_{-}\circ\chib}
  }{
    \nhat_{-}\circ\chib
  }
  =0 .
\end{align}

All in all, we may take Section~\ref{sec:var_prin_solid_decomp}'s action~\eqref{eq:action_variably_rotating} as the starting point for our analysis of the fluid--solid body, always remembering that it is subject to the strongly-imposed reconstruction equation~\eqref{eq:reconstruction_equation} and tangential slip constraints~\eqref{eq:tangential_slip_constraint_motion} and \eqref{eq:tangential_slip_constraint_test_functions}, as well as to the weakly-imposed Euclidean constraints.

\subsection{Exact equations of motion}
\label{sec:fluid_solid_body_eoms_exact}

\subsubsection{Variation of the action}

To derive the fluid--solid Euler--Lagrange equations we may at first vary the action \textit{exactly} as we did in Section~\ref{sec:single_body_eoms}. The necessity to include the tangential slip  constraint on the motion poses no particular problem; given that we will impose it directly on the equations of motion, the algebra required to vary the action is the same as before. Moreover, the forms that we chose for the test-functions in Section~\ref{sec:lag_mult_ident_solid} trivially respect the tangential slip constraint, so it remains the case that the Lagrange multipliers are
\begin{align}
  \alphab &= 0    \\
  \betab  &= \Omb .
\end{align} 
Eqs.~\eqref{eq:eoms_variable_rotation_weak_form_final} are therefore still valid in the present case. We must now impose the tangential slip constraint suitably on the test functions.

\subsubsection{Imposing tangential slip}

On a practical level, we will impose eq.~\eqref{eq:tangential_slip_constraint_test_functions} weakly even as we continue to impose eq.~\eqref{eq:tangential_slip_constraint_motion} directly on the motion. To do that we define new test-functions \( \pilag : \SigFS \times \III \rightarrow \reals \) and add to the equations of motion the term
\begin{align}
\label{eq:FS_constraint_weak_full}
  -
  \intsurf{\SigFS}{
    \pilag
    \ip{
      \ww_{+}\circ\brak{\chib^{-1}}-\ww_{-}
    }{
      J
      \FF^{-T}
      \cdot
      \nhat
    }
  } .
\end{align}
Henceforward, an unsubscripted variable in the vicinity of \( \SigFS \) that could be evaluated on either side of the boundary should be understood to take the value on the lower side; for example, \( \evaluateat{\FF}{\SigFS} = \evaluateat{\FF_{-}}{\SigFS} \). Also, we will loosely refer to \( \pilag \) as a `Lagrange multiplier' because, just like \( \alphab \) and \( \betab \), it is there to enforce a constraint weakly.

We now augment the equations of motion to incorporate the tangential slip constraints on both the motion and test-functions. Nothing changes except in the momentum equation:
\begin{align}
\label{eq:fluid_solid_weak_form_intermediate__momentum}
  &
  \intvol{\BB}{
    \rho
    \ip{
       \dot{\vv}
      +\Omb^{2}\cdot\phib
      +2\Omb\cdot\vv
      +\dot{\Omb}\cdot\phib
    }{
      \ww
    }
  }
  +
  \intvol{\BB}{
    \ip{\TT+\NNN}{D\ww}
  }
  +
  \intsurf{\partial\BB}{
    \ip{\NNN\cdot\nhat}{\ww}
  }
  \nonumber\\
  &\qquad
  -
  \intsurf{\SigFS}{
    \pilag
    \ip{
      \ww_{+}\circ\brak{\chib^{-1}}-\ww_{-}
    }{
      J\FF^{-T}\cdot\nhat
    }
  }
  =0  ,
\end{align}
where we recall that \( \chib \) is defined through \( \phib_{-}\circ\chib = \phib_{+} \).

\subsubsection{Identification of the Lagrange multiplier}

To find the multiplier enforcing tangential slip on the test-functions we integrate eq.~\eqref{eq:fluid_solid_weak_form_intermediate__momentum}'s term in \( D\ww \) by parts. By requiring that the strong-form of the momentum equation hold within each sub-region, and by using the 
traction-free boundary conditions on the surface, we are then left with 
\begin{align}
  &
  -
  \intsurf{\SigFS}{
    \pilag
    \ip{
      \ww_{+}\circ\brak{\chib^{-1}}-\ww_{-}
    }{
      J\FF^{-T}\cdot\nhat
    }
  }
  \nonumber\\
  &\qquad
  +
  \intsurf{\SigFS}{\braksq{
    \ip{\brak{\TT_{-}+\NNN_{-}}\cdot\nhat_{-}}{\ww_{-}}
    +
    \ip{\brak{\TT_{+}+\NNN_{+}}\cdot\nhat_{+}}{\ww_{+}}
  }}
  = 0 .
\end{align}
 Now, \( \nhat_{+} = -\nhat_{-} \) because \( \nhat_{+} \) points outward with respect to the body on the upper side of the boundary. Therefore if we `pair up' the various instances of \( \ww_{-} \) and \( \ww_{+} \) we find
\begin{align}
  &
  \intsurf{\SigFS}{\ip{
    \ww_{-}
  }{
    (\TT_{-}+\NNN_{-})\cdot\nhat + \pilag J\FF^{-T}\cdot\nhat
  }}
  \nonumber\\
  &\qquad
  -
  \intsurf{\SigFS}{\braksq{
    \ip{\ww_{+}}{(\TT_{+}+\NNN_{+})\cdot\nhat}
    +
    \ip{\ww_{+}\circ\brak{\chib^{-1}}}{\pilag J\FF^{-T}\cdot\nhat}
  }}
  = 0.
\end{align}
Because \(\ww_{-}\) and \(\ww_{+}\) are independent, we obtain from the first term
\begin{align}
\label{eq:Pi_is_pressure1}
  (\tt_{-} + \mathbf{n}_{-})  + \pilag J\FF^{-T}\cdot\nhat = 0 ,
\end{align}
where we have defined the traction vectors \(\tt_{-} = \TT_{-}\cdot \nhat\) and \(\mathbf{n}_{-} = \NNN_{-}\cdot \nhat\) for convenience. 
Similarly, from the second part of the equality we find
\begin{align}
\label{eq:Pi_is_pressure2}
  J_{\chib}\,(\tt_{+} + \mathbf{n}_{+})\circ \chib  + \pilag J\FF^{-T}\cdot\nhat = 0 ,
\end{align}
where we have used \( \chib \) to change variables in part of the surface integral, with \(J_{\chib}\) being 
the associated Jacobian. Comparing these equalities, we firstly have
\begin{align}
    \tt_{-} + \mathbf{n}_{-} = J_{\chib}\,(\tt_{+} + \mathbf{n}_{+})\circ \chib. 
\end{align}
This states the equality of mechanical and gravitational tractions at points that are instantaneously adjactent on 
the fluid-solid boundary (note that the surface Jacobian term is due to the tractions being measured 
per unit area on different parts of the reference boundary). From Poisson's equation, it can be show  that 
continuity of the gravitational traction follows automatically \citep[\eg][Section  2.8.3]{dahlen1998theoretical}, and hence this equality reduces to 
\begin{align}
    \tt_{-} = J_{\chib} \tt_{+}\circ \chib, 
\end{align}
which is a familiar result \citep[\eg][]{Woodhouse_1978,dahlen1998theoretical,Al_Attar_2018}.
From eqs.~\eqref{eq:Pi_is_pressure1} and \eqref{eq:Pi_is_pressure2} we can also see that the combined mechanical and gravitational tractions
on either side of the fluid solid boundary take the form expected within (or adjacent to) a fluid, with \(\varpi\) being the pressure. It
is worth commenting that the occurrence of the gravitational tractions within these equations is a direct result of our use of the 
gravitational stress tensor within the equations of motion. If desired, the gravitational stress terms can be converted into 
body forces via integration by parts, with the corresponding tractions then removed.

\subsubsection{Summary} 
\label{sec:fs_eoms_summary}

The full set of equations of motion for a single, variably-rotating, fluid--solid planet are now
\begin{subequations}
\label{eq:fluid_solid_weak_form_final}
\begin{align}
  &
  \dot{\RR} = \RR\Omb  
  \\
  &
  m\dot{\VV} = 0
  \\
  &
  \brak{\partial_{t} + \smallAd{\Omb}}\II\cdot\Omb = 0
  \label{eq:fluid_solid_weak_form_final__euler}\\
  &
  \intvol{\BB}{
    \rho
    \ip{
       \dot{\vv}
      +\Omb^{2}\cdot\phib
      +2\Omb\cdot\vv
      +\dot{\Omb}\cdot\phib
    }{
      \ww
    }
  }
  +
  \intvol{\BB}{
    \ip{\TT+\NNN}{D\ww}
  }
  +
  \intsurf{\partial\BB}{
    \ip{\NNN\cdot\nhat}{\ww}
  }
  \nonumber\\
  &\qquad
  -
  \intsurf{\SigFS}{
    \pilag
    \ip{
      \ww_{+}\circ\brak{\chib^{-1}}-\ww_{-}
    }{
      J
      \FF^{-T}
      \cdot
      \nhat
    }
  }
  =0 
  \label{eq:fluid_solid_weak_form_final__mom}\\
  &
  \intvol{\BB}{\rho\potweak}
  +
  \frac{1}{4\pi G}
  \intvol{\RThree}{
    \ip{\aa\cdot\nabla\zeta}{\nabla\potweak}
  }
  = 0
  \label{eq:fluid_solid_weak_form_final__poisson}\\
  &
  \intvol{\BB}{\rho\phib} = 0 
  \label{eq:fluid_solid_weak_form_final__trans_cons}\\
  &
  \intvol{\BB}{\rho\vv\wedge\phib} = 0
  \label{eq:fluid_solid_weak_form_final__rot_cons},
\end{align}
\end{subequations}
where we recall that \(\chib:\III \rightarrow \DIFF{\SigFS}\) is defined through the relation  \(\phib_{-}\circ\chib = \phib_{+}\).
After having solved for the Lagrange multipliers these equations are invariant under any transformation of the form
\begin{align}
\label{eq:w_invar_tmp_1893}
  \ww \rightarrow \ww + \aa + \thetab\cdot\phib
\end{align}
for arbitrary \( \aa \) and \(\thetab \) (\cf Section~\ref{sec:lag_mult_ident_solid}).
We are therefore free to constrain the test-functions to satisfy
\begin{subequations}
\label{eq:test_function_constraints__exact}
\begin{align}
  &\intvol{\BB}{\rho\ww} = 0 \\
  &\intvol{\BB}{\rho\ww\wedge\phib} = 0 .
\end{align}
\end{subequations}
Such constraints will be useful within certain applications. In particular, they will lead in Section~\ref{sec:fluid_solid_linearisation} to the test-functions satisfying identical constraints to the linearised motion; this being necessary within Galerkin methods. 

To summarise, we have obtained equations of motion that describe the deformation of a rotating, self-gravitating, fluid--solid elastic body in terms of:
\begin{enumerate}
	\item the motion of its centre of mass, \( \Phib \);
	\item a rotation about that centre of mass, \( \RR \);
	\item and elastic internal motion \( \phib \) with respect to the frame defined by \( \Phib \) and \( \RR \).
\end{enumerate}
Comparing these results to AC18, who discussed variational principles for fluid-solid bodies relative to a steadily rotating reference frame, we see that no additional complications arise from the use of the Tisserand frame.

\subsection{The fluid relabelling group}
\label{sec:vorticity_maybe}

Recall from Section~\ref{sec:single_body_particle_relabelling} that the action of a solid elastic body is form-invariant under particle-relabelling transformations. Such transformations
do not, in general, constitute a full symmetry of the action because they change the 
material parameters \((\rho,W)\). It is interesting to ask whether there exist 
non-trivial transformations, \(\xib \in \DIFF{}\), such that these parameters are invariant, 
this meaning 
\begin{subequations}
    \begin{align}
        \rho \cdot \xib = \rho
    \end{align}
    \begin{align}
        W \cdot \xib = W,
    \end{align}
\end{subequations}
where we recall eq.~\eqref{eq:prt_rho_W_defn__rho} and \eqref{eq:prt_rho_W_defn__W} for 
the definition of these group actions. 

To examine this question, we start with 
invariance of density. For an arbitrary subset \(U \subseteq \BB\), we have
\begin{align}
    \intvol{U}{\rho} = \intvol{\xib(U)}{J_{\xib} \rho \circ \xib}, 
\end{align}
where we have used \(\xib\) to change variables in obtaining the right hand side of the equality. 
Supposing that \(\xib\) leaves \(\rho\) invariant, we then obtain the relation
\begin{align}
    \intvol{U}{\rho} = \intvol{\xib(U)}{\rho}.
\end{align}
While such an equality cannot hold in general, there are situations where it does. For example, 
when \(\rho\) is a constant and \(\xib\) is volume preserving. More generally, it is not difficult to see that the stated equality 
holds if there exists a reference configuration of the body with respect to which the 
density it constant.  In physical terms, this situation
describes a body that has  a uniform composition, meaning
 that spatial variations in density within the chosen reference configuration are due solely to 
deformation from this hypothetical undeformed state. In fact, the existence of 
such a reference configuration follows from a theorem of \cite{Moser_1965}.
These arguments extend to piecewise continuous densities so long 
as the support of \(\xib\) is suitably limited. We conclude that
there  do exist non-trivial particle relabelling transformations that 
leave the density invariant. 

Turning to the invariance of the strain energy function we can apply a similar method. 
Again for arbitrary \(U \subseteq \BB\), we can use \(\xib\) to change variables to 
arrive at
\begin{align}
    \intvol{U}{W[\xx,\FF(\xx)]} = \intvol{\xib(U)}{J_{\xib}(\xx) W[\xib(\xx),(\FF\circ\xib)(\xx)]}, 
\end{align}
where \(\FF\) is here an arbitrary field taking values in \(\GL(3)\). If \(W\) is invariant
under \(\xib\), then this equality reduces to
\begin{align}
    \intvol{U}{W[\xx,\FF(\xx)]} = \intvol{\xib(U)}{W[\xx,(\FF\circ\xib)(\xx) \FF_{\xib}(\xx)  ]}, 
\end{align}
and because \(\FF\) is arbitrary we are free to choose \(\FF  = \id\). 
In this manner we have arrived at the identity
\begin{align}
    \intvol{U}{W(\xx,\id)} = \intvol{\xib(U)}{W[\xx, \FF_{\xib}(\xx)  ]}, 
\end{align}
that must hold if the action of \(\xib\) leaves the strain energy function invariant. 
This identity has a simple physical interpretation. On the left hand side we have
the elastic potential energy of the particles contained within \(U\) within the reference 
configuration, while on the right is the energy of these particles following their deformation 
under \(\xib\). It follows that a particle relabelling transformation leaves \(W\) invariant 
only if this transformation, regarded as a configuration of the body, would be associated with no change in the elastic potential energy. For an elastic solid, this is only possible for elements of the Euclidean group, but these can be discounted because we also require that 
\(\BB\) be mapped onto itself (a rigid rotation of a spherically symmetric reference body about its centre would 
be a counter example, but this is not interesting in practice). It follows that
for solid elastic bodies there are no non-trivial particle relabelling transformations that leave the
strain energy invariant. 

The preceding argument can be specialised to a body containing a fluid sub-region, \(\BB_{F}\), 
with more interesting results. Supposing that \(U \subseteq \BB_{F}\), and recalling eq.~\eqref{eq:fs_strain_energy}, we find that the above necessary identity reduces to
\begin{align}
    \intvol{U}{V(\xx,1)} = \intvol{\xib(U)}{V[\xx, J_{\xib}(\xx)  ]}. 
\end{align}
The interpretation of this equation is as before: that the elastic potential 
energy of the particles within \(U\) is unchanged under a deformation by \(\xib\). 
Here, however, this could hold for non-trivial mappings. Indeed, the 
characteristic property of an elastic fluid is that its potential energy is
determined  by changes to volume and not to shape \citep[\eg][]{Noll_1974}.
To go further, we assume without essential loss of generality, that
the reference configuration is stress free. This means that \(J = 1\)
is a minimum of the \(V\), and we are free to adjust \(V\) such that 
this minimum value is zero. If, further, we suppose
that \(V\) is a non-negative and strictly convex function of its second argument
then the equality requires \(J_{\xib} = 1\). Note that the latter
assumptions are met  for conventional forms of \(V\)
such as a fluid Saint-Venant--Kirchhoff material
\citep[\eg][]{Holzapfel_2000}
for which 
\begin{align}
    V(\xx,J) = \frac{1}{2}\kappa(\xx) (\ln J) ^{2},
\end{align}
with \(\kappa\) a positive definite constant (equal to the bulk modulus within the corresponding linearised  theory). 
We conclude that non-trivial particle relabelling transformations that leave \(V\) invariant exist so long as
they are volume preserving and satisfy
\begin{align}
    V[\xib(\xx),J] = V(\xx,J).
\end{align}
Such conditions can, in general, be met.

Summarising the above arguments, we have shown that the action is invariant 
under a particle relabelling transformation if its support is restricted to 
\(\BB_{F}\), and if the following conditions are met
\begin{subequations}
    \begin{align}
        J_{\xib} = 1
    \end{align}
    \begin{align}
        \rho \circ \xib = \rho
    \end{align}
    \begin{align}
        V[\xib(\xx),J] = V(\xx,J),
    \end{align}
\end{subequations}
where in the final equation \(\xx \in \BB_{F}\) and \(J\) is arbitrary.
These conditions define a sub-group of the particle relabelling transformations that 
we term the \textit{fluid relabelling group} \( \GeoDiff{\BB} \). Such transformations
can be said to map level surfaces of both \(\rho\) and \(V\) into themselves, and hence 
following such a transformation it would appear to an observer shown only the 
initial and final states that nothing had happened. This group is equivalent to the \textit{geostrophic modes} discussed by D\&T (p.~118) within the context of linearised elasticity.

To proceed, we characterise the Lie algebra of \( \GeoDiff{\BB} \). Varying \( \xib \rightarrow \idgen + \delta\xib \), we find quickly that
\begin{align}
  &\mathrm{div}\delta\xib = 0
  \\
  &\ip{D\rho}{\delta\xib} = 0
  \\
  &\ip{D_{\xx}V}{\delta\xib} = 0 .
\end{align}
Thus, the Lie algebra comprises divergence-free vector fields supported
in \(\BB_{F}\) that are everywhere tangent to the  level surfaces of density and the strain energy.
In fact, it would be usual for these level surface to be coincident 
\citep[\eg][]{Stevenson_1987}, 
in which case
we could drop one of the final two conditions. As a special case, there are neutrally 
stratified fluids for which \( \GeoDiff{\BB} \) is equal to all volume preserving 
diffeomorphisms with support contained in \(\BB\).

 Noether's theorem yields the conservation law arising from invariance of the action under \( \GeoDiff{\BB} \). Let
\begin{align}
  \phib_{s} = \phib\cdot\xib_{s} ,
\end{align}
where \( s \mapsto \xib_{s} \) is a curve in \( \GeoDiff{\BB} \) with \(\xib_{0} = \idgen\). We know from earlier that
\( \intvol{\BB}{\ip{D_{\vv}\lag}{\delta\phib}} \)
is conserved, with
\begin{align}
  \delta\phib = \evaluateat{\dbyd{s}\phib_{s}}{s=0} .
\end{align}
For the chosen form of \( \phib_{s} \),
\begin{align}
  \delta\phib = \FF\cdot\delta\xib , 
\end{align}
where we define 
\begin{align}
  \delta\xib = \evaluateat{\dbyd{s}\xib_{s}}{s=0} .
\end{align}
Finally, since \( D_{\vv}\lag = \rho\brak{\vv+\Omb\cdot\phib} \), the conserved quantity is
\begin{align}
  \intvol{\BB}{\rho\ip{\FF^{T}\cdot\brak{\vv+\Omb\cdot\phib}}{\delta\xib}}
\end{align}
for arbitrary \( \delta\xib\) within the Lie algebra of \(\GeoDiff{\BB}\).
The derivation of such a conservation law from a particle relabelling symmetry resembles work within the context of fluid 
mechanics  \citep[\eg][]{Mueller_1995,Albert_1997,Nevir_2004,Bridges_2005}, 
but a precise correspondence has not been established. 

The full implications of this conservation law will be discussed elsewhere. But we conclude this discussion with the case of a neutrally stratified fluid. 
The conservation law now states that the restriction of \(\rho\FF^{T}\cdot\brak{\vv+\Omb\cdot\phib}\) 
to \(\BB_{F}\) is orthogonal to the space of divergence-free vector fields on this domain with vanishing normal components. 
From the Helmholtz decomposition theorem \citep[\eg][]{Abraham_2012}, it follows that for some 
scalar field, \(\chi\) say, we have
\begin{align}
  \vv+\Omb\cdot\phib = \frac{1}{\rho} \FF^{-T}\cdot \nabla \chi.
\end{align}
In this manner we have arrived at a referential form of the potential representation for the velocity vector in neutrally stratified fluid regions that is commonly applied within numerical work
\citep[\eg][]{Komatitsch_2002,Leng_2019}.  Further analysis of this
conservation law may yield analogous representations of the velocity  that are applicable within stratified fluids
\cite[\cf][]{Chaljub_2004}.

\subsection{Accounting for the oceans and other surface loads}
\label{sec:surface_load}

We now consider how the foregoing equations can be modified to incorporate a thin surface layer such as an ocean and/or an icesheet. Let us regard the layer as an `extra' body \( \widetilde{\BB} \) imposed on top of the original reference body \( \BB \). In effect, we rename \( \BB \rightarrow \BB\cup\widetilde{\BB} \). We then integrate the momentum-equation in \( \widetilde{\BB} \) by parts, and impose that the strong-form of the momentum-equation holds within \( \widetilde{\BB} \). This gives the following modified momentum-equation within \( \BB \):
\begin{align}
  &
  \intvol{\BB}{
    \rho
    \ip{
       \dot{\vv}
      +\Omb^{2}\cdot\phib
      +2\Omb\cdot\vv
      +\dot{\Omb}\cdot\phib
    }{
      \ww
    }
  }
  +
  \intvol{\BB}{
    \ip{\TT+\NNN}{D\ww}
  }
  +
  \intsurf{\partial\BB}{
    \ip{\tt + \NNN\cdot\nhat}{\ww}
  }
  \nonumber\\
  &\qquad
  -
  \intsurf{\SigFS}{
    \pilag
    \ip{
      \ww_{+}\circ\brak{\chib^{-1}}-\ww_{-}
    }{
      J
      \FF^{-T}
      \cdot
      \nhat
    }
  }
  =0   ,  
\end{align}
where \( \tt \) is the traction applied by the surface layer. This is an exact result that could be used in modelling, say, the coupled evolution of the solid Earth and an ice-sheet. However, in many situations it is sufficient to approximate \( \tt \) so as to avoid explicit consideration of the surface layer's dynamics. For example, \citet{Komatitsch_2002} use the \textit{water column approximation} to derive a form of \( \tt \) that is useful for describing the ocean's effect on seismic waves within the solid Earth at sufficiently long periods. Another example is discussed below in the context of quasi-static problems. Note that we must also take account of \( \widetilde{\BB} \) in the Poisson equation, MoI and total mass, but that simple thin-layer approximations are sufficient for that purpose.

\subsection{Equilibrium configurations and quasi-static deformation}
\label{sec:equilibrium_configs}

\subsubsection{Equations of motion}
\label{sec:eom_eqm}

Relative to the Tisserand frame, a relative equilibrium for a planet is characterised by \(\VV\), \(\Omb\), \(\phib\), and \(\zeta\) being 
independent of time. Such a planet can still be moving at a constant velocity through inertial space and rotating steadily, 
but there is no internal deformation. In this situation, the weak form of the momentum equation reduces to
\begin{align}
  &
  \intvol{\BB}{
    \rho
    \ip{
       \Omb^{2}\cdot\phib
    }{
      \ww
    }
  }
  +
  \intvol{\BB}{
    \ip{\TT+\NNN}{D\ww}
  }
  +
  \intsurf{\partial\BB}{
    \ip{ \NNN\cdot\nhat}{\ww}
  }
  \nonumber\\
  &\qquad
  -
  \intsurf{\SigFS}{
    \pilag
    \ip{
      \ww_{+}\circ\brak{\chib^{-1}}-\ww_{-}
    }{
      J
      \FF^{-T}
      \cdot
      \nhat
    }
  }
  =0,    
\end{align} 
while the Euler equation becomes
\begin{align}
    \smallAd{\Omb}(\II \cdot \Omb) = 0.
\end{align}
As ever, it is assumed that the tangential slip constraint on the motion has been imposed strongly. We also have the 
translational
constraint in eq.~\eqref{eq:fluid_solid_weak_form_final__trans_cons} on the internal motion, but 
note that the angular momentum constraint in eq.~\eqref{eq:fluid_solid_weak_form_final__rot_cons} holds trivially at equilibrium.

Within this section, we consider quasi-static deformation associated with prescribed variations in a thin surface layer, 
this situation being, for example, relevant to the study of GIA. As discussed above, the presence of such a surface layer
can be incorporated into the momentum equation by prescribing a suitable surface traction, \(\tt\), 
while corresponding modifications are made to other equations (detailed below). 
Within the surface layer, the strong form of the momentum equation reads
\begin{align}
    \rho \Omb^{2} \phib - \Div (\TT + \NNN) = 0, 
\end{align}
or equivalently as
\begin{align}
    \rho \Omb^{2} \phib - \Div \TT - \rho \gam  = 0, 
\end{align}
where we have written \( \gam = - \FF^{-T} \nabla \zeta\) for the referential  gravitational acceleration. To first-order accuracy in the 
layer thickness, we can  integrate this equation to obtain the expression
\begin{align}
    \tt =  \sigma ( \gam -\Omb^{2}\phib),
\end{align}
for the applied traction, where \(\sigma\) denotes a mass per unit area that we will call the \textit{load}.
Using this result, we can modify the quasi-static momentum equation to read
\begin{align}
  &
  \intvol{\BB}{
    \rho
    \ip{
       \Omb^{2}\cdot\phib
    }{
      \ww
    }
  }
  +
  \intvol{\BB}{
    \ip{\TT+\NNN}{D\ww}
  }
  +
  \intsurf{\partial\BB}{
    \ip{ \sigma ( \gam -\Omb^{2}\phib) + \NNN\cdot\nhat}{\ww}
  }
  \nonumber\\
  &\qquad
  -
  \intsurf{\SigFS}{
    \pilag
    \ip{
      \ww_{+}\circ\brak{\chib^{-1}}-\ww_{-}
    }{
      J
      \FF^{-T}
      \cdot
      \nhat
    }
  }
  =0.  
\end{align} 
The load also enters into the weak form of the Poisson equation as
\begin{align}
    \intvol{\BB}{\rho\,\potweak} + 
    \intsurf{\partial \BB}{ \sigma\, \potweak} 
  +
  \frac{1}{4\pi G}
  \intvol{\RThree}{
    \ip{\aa\cdot\nabla\zeta}{\nabla\potweak}} = 0,
\end{align}
into the definition of the MoI tensor (defined relative to the Tisserand frame)
\begin{align}
     \II\cdot\Omb'
  &\equiv 
  \intvol{\BB}{\rho\brak{\Omb'\cdot\phib}\wedge\phib} + \intsurf{\partial \BB}{\sigma\brak{\Omb'\cdot\phib}\wedge\phib},
\end{align}
for arbitrary \(\Omb' \in \soThree\), and finally within the translational constraint
\begin{align}
    \intvol{\BB}{\rho \phib} + \intsurf{\partial \BB}{\sigma \phib} = 0.
\end{align}
In this manner, we have obtained a complete set of equations for modelling quasi-static deformation due to
prescribed variations within a surface layer subject to a thin-sheet approximation. Within these equations, 
deformation is driven by the load, \(\sigma\), which is a given function on \(\partial \BB \times \III\).

\subsubsection{Invariance under the Euclidean group}
 Suppose that \((\Omb, \phib, \zeta)\) solve the quasi-static equations of motion subject to a given load, \(\sigma\).
 For  arbitrary \(\QQ \in \SOThree\), it is readily seen that
 \begin{subequations}
     \begin{align}
         \Omb \mapsto \BigAd{\QQ}\cdot\Omb,
     \end{align}
     \begin{align}
         \phib \mapsto \QQ \cdot \phib,
     \end{align}
     \begin{align}
         \zeta \mapsto \zeta,
     \end{align}
 \end{subequations}
 provides another solution. Indeed, the only non-trivial step is checking that the MoI tensor transforms as
 \begin{align}
     \II \mapsto \BigAd{\QQ} \II \BigAd{\QQ^{T}}, 
 \end{align}
 which follows directly from its definition. Here we have arrived at the familiar fact that 
 solutions of quasi-static problems are defined only up to an arbitrary rotation 
\citep[\eg][]{marsden1994mathematical}. 
The fact that \(\Phib\) drops out of the equilibrium equations shows that the solution is also defined only up to an arbitrary translation.

 There are two ways to address this non-uniqueness. First, one can restrict attention to 
 observations that are invariant under \(\EThree\) such as relative distances. 
 The non-uniqueness is then simply a gauge-freedom within the problem, and can be fixed 
 in whatever manner is convenient. The alternative is to remember that quasi-static 
 problems arise as an approximation to the full dynamics in cases where the forcing 
 is slow relative to the time-scales of internal deformation. We can, therefore, appeal
 to conservation laws from the full problem to fix the undetermined Euclidean
 transformation. To see how this is done suppose for definiteness that initially 
 the planet was at rest relative to inertial space (an inertial frame can always be chosen for which this is true because \(\VV\) is constant). It follows that the net 
 linear momentum is always zero, and hence there can be no translational 
 part to the quasi-static solution. Regarding rotation, it follows from the 
 quasi-static Euler equation that \(\Omb\) is always  parallel to the 
 spatial angular momentum. If, in solving the quasi-static problem, 
 this condition has not been met, then we simply rotate \(\Omb\)
 along with the other variables so that it holds. The necessary rotation 
 is determined uniquely up to a rotation parallel to the angular momentum whose value is determined, if necessary, 
 by appeal to the initial conditions
 on the rotation matrix, \(\RR\), that described the Tisserand frame. 
 
\subsubsection{Invariance under the fluid relabelling group}

Quasi-static calculations in planets possessing fluid sub-regions
present additional challenges \citep[\eg][]{Longman_1963,Dahlen_1974,Wunsch_1974,Crossley_1975_static}.
A satisfactory resolution was provided by \cite{Dahlen_1974}, who observed that linearisation 
of the displacement was not valid within fluid regions. Instead, he presented a mixed formulation
of the problem, with solid regions described referentially, and fluid regions spatially. In particular, 
it was shown that all relevant dynamic variables within fluid regions could be described in terms
of the perturbed gravitational potential. Such an approach provides a practical method
for quasi-static calculations that has been applied widely \citep[\eg][]{Tromp_1999,Al_Attar_2013,Crawford_2018}.
It is not clear whether Dahlen's approach can be extended to non-linear quasi-static calculations. Here, instead,
     we  outline an alternative method that should work for both linearised and 
non-linear calculations. 

Recall from Section~\ref{sec:vorticity_maybe} that the action of an elastic body is 
invariant under the action of the fluid relabelling group, \(\GeoDiff{\BB}\). Within
the context of quasi-static calculations, this implies that solutions can only be defined
up to an element of this infinite-dimensional group. Indeed, this is, precisely the reason that linearisation of the displacement in 
fluid regions cannot be justified, as Dahlen described
in different language. Crucially however, we know from our earlier discussion that
the action of \(\GeoDiff{\BB}\) leaves \(\phib\) and \(\zeta\) unchanged outside \(\BB_{F}\), 
and has no effect on \(\Omb\). From an observational perspective, therefore, this is a gauge symmetry,
with any solution of the problem being as good as another. There remains, however, the practical question
as to how \textit{any} solution can be obtained numerically in a well-posed manner. To proceed, we 
can seek the solution of the quasi-static equations that minimises the following functional
\begin{align}
  \frac{1}{2}\intvol{\BB_{F}}{\mu \|\CC\|^{2}} ,
\end{align}
where \(\mu\) is a positive-definite scalar field. Note that this functional is, by design, invariant under
\(\SOThree\), and so our earlier discussion of that group remains valid. In essence, this choice
amounts to picking from amongst all possible solutions the one that is closest to a Euclidean 
transformation within fluid regions. Other functionals could be chosen, with all the matters
being that they lead to a well-posed problem. To implement this idea practically, a Lagrange multiplier 
method could be applied \citep[\cf][]{Seliger_1968,Al_Attar_2010}. A full discussion of this idea
will be presented elsewhere, with its practical viability to be established.

\subsection{Linearisation}
\label{sec:fluid_solid_linearisation}

If a planet at equilibrium is disturbed by some small force, then we can approximate its motion by linearising about that equilibrium state. For definiteness, we imagine that a small \textit{stress-glut} sets the planet in motion, so we introduce a dimensionless perturbation parameter \( \epsilon \) and write
\begin{align}
  W(\xx,t,\FF) = U(\xx,\CC) -\frac{1}{2}\epsilon\ip{\stressGlut(\xx,t)}{\CC} .
\end{align}
To linearise about the equilibrium described in Section~\ref{sec:eom_eqm} we substitute the perturbative \textit{ansatz}
\begin{subequations}
\begin{align}
  \phib &= \phib\z + \epsilon\uu\f   \\
  \chib &= \idgen  + \epsilon\chib\f \\ 
  \RR   &= \RR\z\braksq{\id+\epsilon\thetab\f} \\
  \Omb  &= \Omb\z + \epsilon\mm\f \\
  \zeta &= \zeta\z + \epsilon\zeta\f 
\end{align}
\end{subequations}
into the EoMs \eqref{eq:fluid_solid_weak_form_final}. 
In principle we could perturb the test-functions too. After all, the constraints cause \( \ww \) and \( \pilag \) to intertwine closely with \( \phib \). However, it is easy to show that the \( \order\brak{\epsilon} \) part of the test-functions would simply repeat the zeroth-order equations of motion. Therefore we do not perturb \( \ww \), \( \pilag \) or \( \chi \). For the moment we also ignore the CoM position \( \Phib \) because it decouples fully from the other variables.

\subsubsection{Derivation of the linearised equations}

First we linearise the Euclidean constraints. Given that we take the zeroth-order motion to satisfy the equilibrium equations
we find readily that
\begin{subequations}
\label{eq:euclidean_constraints_linear}
\begin{align}
  &\intvol{\BB}{\rho\uu\f} = 0
  \label{eq:euclidean_constraints_linear__u}\\
  &\intvol{\BB}{\rho\dot{\uu}\f\wedge\phib\z} 
  =
  \dbyd{t}\intvol{\BB}{\rho\uu\f\wedge\phib\z} = 0 ,
  \label{eq:euclidean_constraints_linear__j}
\end{align}
\end{subequations}
\begin{align}
\label{eq:euclidean_constraints_zeroth}
  \intvol{\BB}{\rho\phib\z} = 0 .
\end{align}
We also recall that with \( \RR\z \) having been set at equilibrium, the initial condition \( \thetab\f(0) \) constitutes a gauge-freedom of the first-order problem; we set
\begin{align}
\label{eq:constraint_new_ang_mom}
  \intvol{\BB}{\rho\uu\f\wedge\phib\z} = 0 
\end{align}
by choosing \( \thetab\f(0) \) suitably.

Next, the reconstruction equation reduces to
\begin{align}
\label{eq:nmc_linearised_weak_form__main_body__rec}
  \brak{\partial_{t} + \smallAd{\Omb\z}}\cdot\thetab\f = \mm\f ,
\end{align}
while the Euler equation becomes
\begin{align}
\label{eq:nmc_linearised_weak_form__main_body__euler}
  \II\z\cdot\dot{\mm}\f 
  + 
  \smallAd{\Omb\z}\brak{\II\z-\lambda}\cdot\mm\f 
  +
  \brak{\partial_{t} + \smallAd{\Omb\z}}\II\f\cdot\Omb\z
  = 0
\end{align}
upon using the equilibrium solution \( \II\z\cdot\Omb\z = \lambda\Omb\z \). Linearising the momentum and Poisson equations involves tedious algebra that we have consigned to Appendix~\ref{sec:app:lin}. We find the Poisson equation
\begin{align}
  \intvol{\BB}{\ip{
    D\uu\f
  }{
    \Gamt\cdot\nabla\potweak
  }}
  +
  \frac{1}{4\pi G}
  \intvol{\RThree}{
    \ip{\aa\z\cdot\nabla\zeta\f}{\nabla\potweak}
  }
  = 0
\end{align}
and the momentum equation
\begin{align}
\label{eq:nmc_linearised_weak_form__main_body__momentum}
  &
  \intvol{\BB}{
    \rho
    \ip{
       \ddot{\uu}\f 
      +{\Omb\z}^{2}\cdot\uu\f 
      +2{\Omb\z}\cdot\dot{\uu}\f 
    }{
      \ww
    }
  }
  +
  \intvol{\BB}{\ip{
    \brak{\AAA +\Gam}\cdot D\uu\f 
  }{
    D\ww
  }}
  \nonumber\\
  &\quad 
  +
  \intsurf{\partial\BB}{\ip{
    \brak{\Gam\cdot D\uu\f}\cdot\nhat 
  }{
    \ww
  }}
  \nonumber\\
  &\quad 
  +
  \intvol{\BB}{
    \rho
    \ip{
      \brak{
       \dot{\mm}\f + \mm\f \Omb\z + \Omb\z \mm\f 
      }
      \cdot
      \phib\z 
    }{
      \ww
    }
  }
  \nonumber\\
  &\quad
  +
  \intsurf{\SigFS}{
    \pilag
    \ip{D\ww_{+}}{\QFS\cdot\jump{\uu\f}}
  }
  +
  \intsurf{\SigFS}{
    \pilag
    \ip{D\uu\f_{-}}{\QFS\cdot\jump{\ww}}
  }
  \nonumber\\
  &\quad
  +
  \intvol{\BB}{
    \ip{\Gamt\cdot\nabla\zeta\f}{D\ww}
  }
  +
  \intsurf{\partial\BB}{
    \ip{\brak{\Gamt\cdot\nabla\zeta\f}\cdot\nhat}{\ww}
  }
  =
  \intvol{\BB}{\ip{-\FF\z\stressGlut}{D\ww}} ,
\end{align}
with the \textit{elastic tensor} defined as
\begin{align}
  \AAA(\xx) \equiv D^{2}_{F}U\braksq{\xx,\CC\z(\xx)}
\end{align}
and where
\( \QFS \) is defined in eq.~\eqref{eq:QFS_defn},
\( \Gamt \) in eq.~\eqref{eq:Gamt_defn}
and \( \Gam \) in eq.~\eqref{eq:Gam_defn}.
The linearised motion \( \uu\f \) and test-functions \( \ww \) are elements of the function-space \( \fullConsSpace{\phib\z} \) defined by eq.~\eqref{eq:lin_func_space_defn}.

Although the boundary terms have been formulated differently, these equations are equivalent to A18's eq.~(120) aside from our use of the Tisserand frame. The latter equations are equivalent in turn to those of \citet{Woodhouse_1978} when the equilibrium configuration is the identity mapping.

\subsubsection{A streamlined notation}

As in A18, it will be helpful to establish a more streamlined notation before proceeding. For a start, we drop superscript ones and zeroes, and rename \( \phib\z \rightarrow \phibzsh \). Then we define the bilinear forms
\begin{align}
  \Mlin\brak{\uu,\ww}
  &\equiv
  \intvol{\BB}{\rho\ip{\uu}{\ww}}
  \\
  \Wlin\brak{\uu,\ww}
  &\equiv 
  \intvol{\BB}{2\rho\ip{\Omb\cdot\uu}{\ww}}
  \\
  \Hlin\brak{\uu,\ww}
  &\equiv
  \intvol{\BB}{\brakbr{
    \rho
    \ip{
      \Omb^{2}\cdot\uu
    }{
      \ww
    }
    +
    \ip{\brak{\AAA+\Gam}\cdot D\uu}{ D\ww}
  }}
  \nonumber\\
  &\qquad
  +
  \intsurf{\partial\BB}{
    \ip{\brak{\Gam\cdot D\uu}\cdot\nhat}{\ww}
  }
  +
  \intsurf{\SigFS}{
    \pilag
    \braksq{
      \ip{D\ww_{+}}{\QQ\cdot\jump{\uu}}
      +
      \ip{D\uu_{-}}{\QQ\cdot\jump{\ww}}
    }
  }.
\end{align}
For the linearised rotational parts of the momentum equation we may also usefully define
\begin{align}
  \Bnmc_{1}\brak{\omb,\ww}
  &\equiv
  \intvol{\BB}{
    \rho
    \ip{
      \omb\cdot\phibzsh
    }{
      \ww          
    }
  }
  \\
  \Bnmc_{2}\brak{\omb,\ww}
  &\equiv
  \intvol{\BB}{
    \rho
    \ip{
      \brak{\omb\Omb + \Omb\omb}\cdot\phibzsh
    }{
      \ww         
    }
  } .
\end{align}
For the Euler equation, on the other hand, we need
\begin{align}
  \Clin\brak{\uu}
  &\equiv 
  \II\f\cdot\Omb
  =
  \intvol{\BB}{\rho\braksq{
    \brak{\Omb\cdot\phibzsh}\wedge\uu
    +
    \brak{\Omb\cdot\uu}\wedge\phibzsh
  }} .
\end{align}
Next we define the bilinear form
\begin{align}
  \Plin\brak{\zeta,\chi}
  &\equiv
  \frac{1}{4\pi G}
  \intvol{\RThree}{
    \ip{\aa\cdot\nabla\zeta}{\nabla\potweak}
  } 
\end{align}
for the Poisson term, while the terms that couple the motion to the gravity require
\begin{align}
  \Glin\brak{\zeta,\ww}
  &\equiv 
  \intvol{\BB}{
    \ip{\Gamt\cdot\nabla\zeta}{D\ww}
  }
  +
  \intsurf{\partial\BB}{
    \ip{\brak{\Gamt\cdot\nabla\zeta}\cdot\nhat}{\ww}
  } .
\end{align}
With this we can rewrite the linearised equations more conveniently as
\begin{subequations}
\label{eq:nmc_linearised_weak_form__streamlined}
\begin{align}
  &
  \brak{\partial_{t} + \smallAd{\Omb}}\cdot\thetab = \mm
  \\
  &
  \braksq{\II\partial_{t} + \smallAd{\Omb}\brak{\II-\lambda}}\cdot\mm
  +
  \brak{\partial_{t} + \smallAd{\Omb}}\cdot\Clin\brak{\uu} = 0
  \label{eq:nmc_linearised_weak_form__streamlined__euler}
  \\
  &\Mlin\brak{\ddot{\uu},\ww}
  +\Wlin\brak{\dot{\uu},\ww}
  +\Hlin\brak{\uu,\ww}
  +\Glin\brak{\zeta,\ww}
  +\Blin_{1}\brak{\dot{\mm},\ww}  
  +\Blin_{2}\brak{\mm,\ww}  
  =
  \intvol{\BB}{\ip{
    -\FF\stressGlut
  }{
    D\ww
  }}
  \label{eq:nmc_linearised_weak_form__streamlined__momentum}
  \\
  &\Glin\brak{\chi,\uu}
  +\Plin\brak{\zeta,\chi}
  =0
  \label{eq:nmc_linearised_weak_form__streamlined__grav}
\end{align}
\end{subequations}
where \( \uu,\ww \in \fullConsSpace{\phibzsh} \).
\subsection{An application: generalised normal-mode coupling for a rotating \( N \)-layer fluid--solid planet}
\label{sec:nmc1_gen}

To derive the mode-coupling equations we now expand the first-order fields and test-functions on a suitable set of basis functions. For NMC calculations those basis functions are essentially the eigenfunctions of a reference, spherically-symmetric Earth model, but altered so that they satisfy the strong-form constraints imposed on the problem. 
In this paper we will not dwell on the details of imposing the constraints numerically, deferring such considerations to future work on applications. For now we just expand on some set of basis-functions \( \{\hat{\uu}_{n}\} \) and \( \{\hat{\zeta}_{p}\} \) , with the resulting equations assumed to incorporate the necessary constraints. The discussion is applicable to any Galerkin method.

We express the linearised motion and test-functions as
\begin{align}
  \uu(\xx,t) &= \sum_{n}u_{n}(t)\hat{\uu}_{n}(\xx) \\
  \ww(\xx) &= \sum_{n}w_{n}\hat{\uu}_{n}(\xx) ,
\end{align}
for the gravity we set
\begin{align}
  \zeta(\xx,t) &= \sum_{p}z_{p}(t)\hat{\zeta}_{p}(\xx) \\
  \chi (\xx)   &= \sum_{p}c_{p}\hat{\zeta}_{p}(\xx) ,
\end{align}
then we substitute these expansions into eqs.~(\ref{eq:nmc_linearised_weak_form__streamlined}).
Next we take each of the \( w_{n} \) and \( c_{p} \) to be unity in turn -- with all of the others vanishing -- thus projecting out the infinite-dimensional linear-algebraic ODEs
\begin{subequations}
\label{eq:gnmc_equations}
\begin{align}
  &
  \brak{\partial_{t}+\smallAd{\Omb}}\cdot\thetab 
  = 
  \mm
  \\
  &
  \brak{\partial_{t}+\smallAd{\Omb}}
  \tilde{\Cnmcb}\unmcb
  +
  \braksq{\II\partial_{t} + \smallAd{\Omb}\brak{\II-\lambda}}\cdot\mm
  = 0
  \\
  &
   \Mnmcb\ddot{\unmcb} 
  +\Wnmcb\dot{\unmcb} 
  +\Hnmcb\unmcb 
  +\Gnmcb\znmcb  
  +\Bnmcb_{1}\dot{\mm}
  +\Bnmcb_{2}\mm 
  = 
  \fnmcb
  \\
  &
   \Gnmcb^{T}\unmcb
  +\Pnmcb\znmcb
  =
  \zerobsf ,
\end{align}
\end{subequations}
with the various matrices defined by:
\begin{align}
  \Mnmc_{mn}
  &\equiv
  \Mlin\brak{\hat{\uu}_{n},\hat{\uu}_{m}}
\end{align}
and similarly for \( \Wnmc_{mn} \) and \( \Hnmc_{mn} \);
\begin{align}
  \fnmc_{m}
  &\equiv
  \intvol{\BB}{\ip{-\FF\stressGlut}{D\hat{\uu}_{m}}}
\end{align}
for the stress-glut forcing;
\begin{align}
  \braksq{\Bnmcb_{i}\omb}_{m}
  &\equiv
  \Blin_{i}\brak{\omb,\hat{\uu}_{m}}
\end{align}
for arbitrary \( \omb\in\sothree \);
\begin{align}
  \tilde{\Cnmcb}\unmcb 
  &\equiv
  \sum_{n}\Clin\brak{\hat{\uu}_{n}}u_{n} ;
\end{align}
and
\begin{align}
  \Gnmc_{mp}
  &\equiv
  \Glin\brak{\hat{\zeta}_{p},\hat{\uu}_m}
  \\
  \Pnmc_{pq}
  &\equiv
  \Plin\brak{\hat{\zeta}_{q},\hat{\zeta}_{p}}
\end{align}
for the gravity-related terms.
Equations~\eqref{eq:gnmc_equations} may be compared with A18's equations (142-146). We may also Fourier-transform in time and define the frequency-dependent operators
\begin{subequations}
\begin{align}
  \Anmcb
  &\equiv 
  -\nu^{2}\Mnmcb
  +\ii\nu \Wnmcb
  +\Hnmcb
  \\
  \Bnmcb
  &\equiv
  \ii\nu\Bnmcb_{1}+\Bnmcb_{2}
  \\
  \Cnmcb
  &\equiv
  \brak{\ii\nu + \smallAd{\Omb}}\tilde{\Cnmcb}
  \\
  \Dnmcb 
  &\equiv
  \brak{\ii\nu + \smallAd{\Omb}}\II - \lambda\smallAd{\Omb}
  =
  \ii\nu\II + \smallAd{\Omb}\brak{\II - \lambda}
\end{align}
\end{subequations}
to give the following frequency-domain equations of GNMC:
\begin{subequations}
\label{eq:gnmc_equations_freq_dom}
\begin{align}
  \brak{\ii\nu + \smallAd{\Omb}}\thetab &= \mm 
  \label{eq:gnmc_equations_freq_dom__recons} \\
  \Cnmcb\unmcb + \Dnmcb\mm &= \zerob
  \label{eq:gnmc_equations_freq_dom__euler}\\
  \Anmcb\unmcb + \Bnmcb\mm + \Gnmcb\znmcb &= \fnmcb
  \label{eq:gnmc_equations_freq_dom__momentum}\\
  \Gnmcb^{T}\unmcb + \Pnmcb\znmcb &=  \zerobsf 
  \label{eq:gnmc_equations_freq_dom__grav}
\end{align}
\end{subequations}
These equations, together with their time-domain counterparts \eqref{eq:gnmc_equations}, are one of this paper's main results.

\section{Interactions between multiple bodies}
\label{sec:multiple}

In some problems one will need to consider multiple interacting bodies. Such problems can be approached using the formalism we have just presented, as we now show by discussing two problems: one of orbital dynamics (Section~\ref{sec:orbital_dynamics}), the other related to Earth's rotational variations (Section~\ref{sec:prec_nut_trans}). This section is by no means intended to be exhaustive. The examples we give are for methodological illustration -- and even speculation -- so we mainly present actions and comment on physics with minimal mathematical formality. 
We plan to revisit both of the following problems in future application-based work, at which time we will develop these ideas more fully.

\subsection{Application: generalised orbital dynamics}
\label{sec:orbital_dynamics}


Let two fluid--solid planets 
interact gravitationally. The action describing such a system is the sum of the planets' respective `single-body actions' (eq.~\ref{eq:action_variably_rotating}) but with an extra term added to account for the planets' gravitational binding energy. That binding energy is just the spatial gravitational potential of one body integrated against the other's density, so if we define
\begin{align}
  \Gamma(\yy) = \frac{G}{\|\yy\|} ,
\end{align} 
and recall definition~\eqref{eq:internal_elastic_action_defn} of the internal elastic action \( \actionSmall \), then the total action is
\begin{align}
\label{eq:action_kepler}
  \action
  &= 
  \inttime{\III}{\frac{1}{2}m_{1}\|\VV_{1}\|^{2}}
  +
  \inttime{\III}{\frac{1}{2}m_{2}\|\VV_{2}\|^{2}}
  -
  \inttime{\III}{\intvol{\BB_{1}}{\rho_{1}\intvolprime{\BB_{2}}{\rho_{2}'
    \Gamma\braksq{
      \brak{\Phib_{1}+\RR_{1}\cdot\phib_{1}} 
      - 
      \brak{\Phib_{2}+\RR_{2}\cdot\phib_{2}'}
    }
  }}}
  \nonumber\\
  &\qquad\qquad
  +
  \inttime{\III}{\frac{1}{2}\ip{\Omb_{1}}{\II_{1}\cdot\Omb_{1}}}
  +
  \inttime{\III}{\frac{1}{2}\ip{\Omb_{2}}{\II_{2}\cdot\Omb_{2}}}
  +
  \actionSmall_{1}
  +
  \actionSmall_{2}
  .
\end{align}
We have added subscripts in an obvious way. Each \( \phib_{i} \) of course satisfies the tangential-slip constraint~\eqref{eq:tangential_slip_ur} while \( \Omb_{i} \) is related to \( \RR_{i} \) by the reconstruction equation~\eqref{eq:reconstruction_equation}.

Now let us make a change of variables learned from the Keplerian two-body problem. We repackage \( m_{1} \) and \( m_{2} \) into a total mass \( M \) and a \textit{reduced mass} \( \mu \),
\begin{subequations}
\begin{align}
  M        &\equiv m_{1} + m_{2}
  \\
  \mu      &\equiv \frac{m_{1}m_{2}}{M} ,
\end{align}
\end{subequations}
then we define the bodies' relative position \( \rr \) and the \textit{barycentre position} \( \bary \) by
\begin{subequations}
\label{eq:kep_variable_change}
\begin{align}
  \rr(t)   &\equiv \Phib_{1}(t) - \Phib_{2}(t) 
  \\
  \bary(t) &\equiv \frac{1}{M}\braksq{m_{1}\Phib_{1}(t) + m_{2}\Phib_{2}(t)} .
\end{align}
\end{subequations}
These transformations hit the top line of eq.~\eqref{eq:action_kepler} to give
\begin{align}
\label{eq:action_kepler_2}
  \action
  &= 
  \inttime{\III}{\frac{1}{2}M\|\dot{\bary}\|^{2}}
  +
  \inttime{\III}{\frac{1}{2}\mu\|\dot{\rr}\|^{2}}
  -
  \inttime{\III}{\intvol{\BB_{1}}{\rho_{1}\intvolprime{\BB_{2}}{\rho_{2}'
    \Gamma\brak{\rr + \RR_{1}\cdot\phib_{1} - \RR_{2}\cdot\phib_{2}'}
  }}}
  \nonumber\\
  &\qquad
  +
  \brak{\text{`single-body' terms}} .
\end{align}
We see immediately that \( \bary \) is a cyclic variable, \ie the two-body system's overall CoM moves at constant velocity. \( \rr \) on the other hand satisfies
\begin{align}
\label{eq:kepler_eqn_r}
  \mu\ddot{\rr} 
  =
  -
  \intvol{\BB_{1}}{\rho_{1}\intvolprime{\BB_{2}}{\rho_{2}'
    D\Gamma\brak{\rr + \RR_{1}\cdot\phib_{1} - \RR_{2}\cdot\phib_{2}'} 
  }},
\end{align}
as follows quickly by varying action~\eqref{eq:action_kepler_2}. Since \( \rr \) will generally have much greater magnitude than \( \phib_{1,2} \)
 this can clearly be approximated as
\begin{align}
  \ddot{\rr}
  =
  -\frac{GM}{\norm{\rr}^{3}}\rr
  +\order\brak{\frac{\|\phib_{1,2}\|^{2}}{\|\rr\|^{2}}} .
\end{align}
The planets' relative translational motion thus satisfies a modified Keplerian two-body problem. Since \( \rr \) couples to \( \phib_{1,2} \) and \( \RR_{1,2} \), equation~\eqref{eq:kepler_eqn_r} exhibits in principle both spin--orbit coupling and coupling between translation and elastic deformation. 
This is applicable to studies of long-timescale orbital evolution accounting for tidal dissipation \citep{Cuk_2016,Lock_2017}.

As for the elastic motions \( \phib_{1} \) and \( \phib_{2} \), not only are they coupled to \( \rr \), but also, in principle, to each other. Still, body 2's elastic deformation should generally have a negligible effect on body 1's.
This approximation would be relevant to present-day tidal deformation on Earth.
In other words, in order to compute \( \phib_{1} \) one could reasonably approximate the binding energy as
\begin{align}
  \intvol{\BB_{1}}{\rho_{1}\intvolprime{\BB_{2}}{\rho_{2}'
    \Gamma\brak{\rr+\RR_{1}\cdot\phib_{1}}
  }} .
\end{align}
Noting that the integration in \( \xx' \) trivially yields body 2's mass, and rearranging \( \Gamma \)'s argument, it follows that \( \phib_{1} \)'s dynamics would be described by the action
\begin{align}
  \action 
  =
  \actionSmall_{1}
  +
  \inttime{\III}{\frac{1}{2}\ip{\Ombr_{1}}{\IIr_{1}\cdot\Ombr_{1}}}
  -
  m_{2}
  \inttime{\III}{\intvol{\BB_{1}}{\rho_{1}
    \Gamma\brak{\phib_{1} + \RR_{1}^{T}\cdot\rr}
  }}
  +
  \inttime{\III}{\frac{1}{2}\mu\|\dot{\rr}\|^{2}} .
\end{align}
The third term, through which \( \phib_{1} \) feels body 2, represents 
a \textit{tidal potential}. 
That term will also provide a torque for \( \Omb_{1} \) to respond to. The extension to \( N \)-body systems is clear.


\subsection{Application: wobbles, nutations and internal translations of a rotating \( N \)-layer fluid--solid planet}
\label{sec:prec_nut_trans}

It is sometimes of interest to regard each layer of a fluid--solid planet as an individual elastic body; this leads to a problem of interacting bodies that is similar in spirit to what we discussed in Section~\ref{sec:orbital_dynamics}. Such a viewpoint is often natural when studying Earth's \textit{precession and nutation} -- \ie the small, quasi-rigid relative rotations of the Earth's layers \citep[\eg][]{dehant2015precession} -- as well as quasi-rigid relative translations like the \textit{Slichter modes} 
\citep[\eg][]{Slichter_1961,Smith_1976}.
It motivates a more general decomposition of the motion whereby each layer has its own Euclidean motion.

We start with eqs.~\eqref{eq:fluid_solid_weak_form_final} describing an \( N \)-layer fluid--solid planet. Then we regard the restrictions \( \evaluateat{\phib}{\BB_{i}} \equiv \phib_{i} \) as distinct motions and decompose each of them in the same way as we decomposed the overall motion (\cf eqs.~\ref{eq:decomp_fs}):
\begin{align}
\label{eq:internal_internal_decomp}
  \phib_{i} = \Phib_{i} + \RR_{i}\cdot\phibr_{i} .
\end{align}
Recall that the \( \phib \) of eqs.~\eqref{eq:fluid_solid_weak_form_final} really represents internal motion with respect to the Tisserand frame, so in expression~\eqref{eq:internal_internal_decomp} \( \Phib_{i} \) and \( \RR_{i} \) represent \textit{relative Euclidean} motions, and \( \phibr_{i} \) \textit{relative internal} motions. The frame defined by each of the \( \Phib_{i} , \RR_{i} \) is defined by imposing the constraints
\begin{subequations}
\label{eq:internal_internal_constraints}
\begin{align}
  &\intvol{\BB_{i}}{\rho_{i}\phibr_{i}} = 0
  \\
  &\intvol{\BB_{i}}{\rho_{i}\vvr_{i}\wedge\phibr_{i}} = 0 ,
\end{align}
\end{subequations}
once again in precise analogy with eqs.~\eqref{eq:decomp_fs}. Within the original Tisserand frame, we are in effect defining `relative Tisserand frames', one for each layer. Recalling from eq.~\eqref{eq:test_function_constraints__exact} that we constrained the test-functions analogously to the motion, we make a corresponding decomposition of the test-functions:
\begin{subequations}
\label{eq:internal_internal_constraints_test}
\begin{align}
  &\ww_{i} = \Psib_{i} + \RR_{i}\cdot\brak{\wwr_{i}+\Thetabr_{i}\cdot\phibr_{i}}
  \\
  &\intvol{\BB_{i}}{\rho_{i}\wwr_{i}} = 0
  \\
  &\intvol{\BB_{i}}{\rho_{i}\wwr_{i}\wedge\phibr_{i}} = 0 ,
\end{align}
\end{subequations}
where each \( \wwr_{i} \) is an arbitrary test-function with support on \( \BB_{i} \) while \( \Psib_{i} \) and \( \Thetabr_{i} \) are arbitrary elements of \( \RThree \) and \( \soThree \) respectively.
With decompositions~\eqref{eq:internal_internal_decomp} and their constraints~\eqref{eq:internal_internal_constraints}, decomposition~\eqref{eq:decomp_fs}'s original Euclidean constraints -- which of course define the Tisserand frame -- reduce to
\begin{align}
  &\sum_{i=1}^{N}m_{i}\Phib_{i} = 0
  \\
  &\sum_{i=1}^{N}m_{i}\VV_{i}\wedge\Phib_{i}
  +\sum_{i=1}^{N}\BigAd{\RR_{i}}\IIr_{i}\cdot\Ombr_{i}
  =0 ,
\end{align}
where \( \Ombr_{i} = \RR_{i}^{T}\dot{\RR}_{i} \) and each layer's \textit{relative MoI} is defined by (\cf~eq.~\ref{eq:moi_defn_int})
\begin{align}
  \IIr_{i}\cdot\Omb'
  =
  \intvol{\BB_{i}}{\rho_{i}\Omb'\cdot\phibr_{i}\wedge\phibr_{i}} .
\end{align}
Similarly for the test functions:
\begin{align}
  &\sum_{i=1}^{N}m_{i}\Psib_{i} = 0
  \\
  &\sum_{i=1}^{N}m_{i}\Psib_{i}\wedge\Phib_{i}
  +\sum_{i=1}^{N}\BigAd{\RR_{i}}\IIr_{i}\cdot\Thetabr_{i}
  =0 .
\end{align}
The tangential-slip constraints mix together the relative Euclidean and internal motions:
\begin{align}
  &
  \Phib_{i}+\RR_{i}\cdot\phib_{i}\circ\chib = \Phib_{i+1}+\RR_{i+1}\cdot\phib_{i+1} .
\end{align}
Now we substitute all these expressions into eqs.~\eqref{eq:fluid_solid_weak_form_final}.
We continue to regard the referential potential as a single field, although we will sometimes use \( \zeta_{i} \) as a shorthand for the restriction \( \evaluateat{\zeta}{\BB_{i}} \).

Substituting these expressions into eqs.~\eqref{eq:fluid_solid_weak_form_final}, and defining the useful quantities
\begin{align}
  \Ombt_{i} \equiv \Ombr_{i} + \BigAd{\RR_{i}^{-1}}\cdot\Omb ,
\end{align}
can be shown to bring us to the following equations of motion.
The overall Tisserand frame evolves according to
\begin{align}
  &
  \dot{\RR} = \RR\Omb  
  \\
  &
  m\dot{\VV} = 0
  \\
  &
  \brak{\partial_{t} + \smallAd{\Omb}}\II\cdot\Omb = 0 ,
\end{align}
showing that it is indeed unaffected by the planet's internal interactions. Meanhwile, the \( N \) relative Tisserand frames each obey
\begin{align}
  &
  \dot{\RR}_{i} = \RR_{i}\Ombr_{i}  
  \\
  &
  \sum_{i}
  \ip{
    \Psib_{i}
  }{
    m_{i}\brak{\partial_{t}+\Omb}^{2}\cdot\Phib_{i}
  }
  -
  {\sum_{i}}'
  \ip{
    \RR_{i}^{T}\cdot\brak{\Psib_{i+1}-\Psib_{i}}
  }{
    \intsurf{\Sigma_{i}}{\tracp_{i}}
  } 
  = 0
  \\
  &
  \sum_{i}
  \ip{
    \Thetabr_{i}
  }{
    \brak{\partial_{t}+\smallAd{\Ombt_{i}}}\IIr_{i}\cdot\Ombt_{i}
  }
  -
  {\sum_{i}}'
  \intsurf{\Sigma_{i}}{\ip{
    \RR_{i}^{T}\RR_{i+1}\Thetabr_{i+1}
    \cdot
    \brak{\phibr_{i+1}\circ\chib^{-1}}
    -
    \Thetabr_{i}
    \cdot
    \phibr_{i}
  }{
    \tracp_{i}
  }}
  = 0 
\end{align}
for arbitrary \(  \Psib_{i}  \) and \( \Thetabr_{i} \). Here we see the boundary tractions contributing net forces and torques that influence the respective layers' CoMs' motions and orientations. As for the relative internal motions, they obey
\begin{align}
  &
  \sum_{i}
  \intvol{\BB_{i}}{
    \ip{
      \wwr_{i}
    }{
      \rho_{i}\brak{\partial_{t}+\Ombt_{i}}^{2}\cdot\phibr_{i}
    }
  }
  +
  \sum_{i}
  \intvol{\BB_{i}}{
    \ip{\TTr_{i}+\NNNr_{i}}{D\wwr_{i}}
  }
  +
  \intsurf{\partial\BB}{
    \ip{\NNNr_{N}\cdot\nhat}{\wwr_{N}}
  }
  \nonumber\\
  &\qquad
  -
  {\sum_{i}}'
  \intsurf{\Sigma_{i}}{
    \ip{
      \RR_{i}^{-1}\RR_{i+1}\cdot\brak{\wwr_{i+1}\circ\chib^{-1}}-\wwr_{i}
    }{
      \tracp_{i}
    }
  }
  =0 ,
\end{align}
and couple to other layers' relative Euclidean motions.
And of course we still have the Poisson equation
\begin{align}
  &
  \intvol{\BB}{\rho\potweak}
  +
  \frac{1}{4\pi G}
  \intvol{\RThree}{
    \ip{\aa\cdot\nabla\zeta}{\nabla\potweak}
  }
  =0  .
\end{align}

\section{Discussion}
\label{sec:conc}

We have derived equations of motion for variably rotating, self-gravitating, (visco)elastic,  fluid--solid planets relative to the 
Tisserand frame. 
The main results are the exact equations presented in Section~\ref{sec:fs_eoms_summary}, their quasi-static form
in Section~\ref{sec:equilibrium_configs}, and the resulting linearised equations of motion in Section~\ref{sec:fluid_solid_linearisation}.
The final form of the equations of motion obtained differs only rather slightly from standard approaches based on a steadily rotating reference frame. There would, therefore, be no substantial cost or difficult in implementing this new formulation, but by doing so the potential range of applicability is extended. Throughout this paper we have emphasised the role of symmetry, and within Section~\ref{sec:vorticity_maybe} used this approach to identify a seemingly new conservation law related to fluid regions. 
Section~\ref{sec:multiple} extended the `Tisserand approach' to problems concerning the interactions of multiple bodies. This paper's new theoretical framework will form the basis for a range of future studies.

\begin{acknowledgments}
MM's work on this paper has been supported by an EPSRC studentship and a CASE award from BP, and more recently by the European Research Council (agreement 833848-UEMHP) under the Horizon 2020 programme. DA has been supported through the Natural Environment Research Council grant numbers NE/V010433/1 and NE/X013804/1.
\end{acknowledgments}

\subsection*{Data availability statement}

There are no data or code relating to this work.

\bibliographystyle{gji}
\bibliography{my_paper_library}

\appendix

\section{Notations and definitions}

\subsection{Groups}
\label{app_sub:groups}

We define \( \gl(n) \), \textit{the general linear group of dimension n}, to be the set of invertible \( n \times n \) matrices under the operation of matrix multiplication. For a general group \( \mathbf{G} \), a \textit{subgroup} \( \mathbf{H} \) of \( \mathbf{G} \) is a subset of the elements of \( \mathbf{G} \) that is itself a group; \( \mathbf{H} \) is described as a \textit{proper subgroup} of \( \mathbf{G} \) if \( \mathbf{H} \neq \mathbf{G} \). With this, we can define \( \sl(n) \), \textit{the special linear group} of \( n \times n \) matrices with \textit{unit} determinant, which is a proper subgroup of \( \gl(n) \). A particularly important proper subgroup of \( \sl(n) \) is \( \SO(n) \), \textit{the n-dimensional special orthogonal group} whose elements are rotation matrices in \( n \) dimensions; within this work we focus on the case \(n = 3\) for obvious reasons.
General references for the material within this section are \cite{holm2009geometric,iserles2009deqn}.

\subsubsection{The rotation group \( \SO(3) \)}
\label{sec:groups_sothree}

\( \so(3) \) is a matrix Lie group  whose elements satisfy
\begin{align}
  \RR^{-1}  &= \RR^{T} \\
  \det{\RR} &= 1 .
\end{align} 
Elements of \( \SOThree \) act on elements of \( \RThree \) from the left according to
\begin{align}
  \aa \mapsto \RR\cdot\aa.
\end{align}
To understand the associated  Lie algebra \( \soThree \), consider a one-parameter family of rotation matrices \( s \mapsto \RR(s) \), with \( \RR(0) = \id \). If we differentiate the relation \( \RR(s)\RR(s)^{T} = \id \), we find that
\begin{align}
  \dot{\RR}\RR^{T} + \RR\dot{\RR}^{T} = 0 ,
\end{align}
which can be rearranged to give
\begin{align}
  \dot{\RR}\RR^{T} = -\brak{\dot{\RR}\RR^{T}}^{T} .
\end{align}
This shows that \( \dot{\RR}\RR^{T} \) is \textit{antisymmetric}. Evaluating about \( s=0 \) gives
\begin{align}
  \evaluateat{\dbyd{s}\RR}{s=0}
  =
  \Omb ,
\end{align} 
for some antisymmetric matrix \( \Omb \). This shows that \( \SOThree \)'s tangent-space at the identity, which is \( \soThree \) by definition, consists of all antisymmetric \( (3 \times 3) \) matrices. 
Such matrices have three independent components, so we may write \( \soThree \)'s left-action on \( \RThree \) as
\begin{align}
  \aa \mapsto \Omb\cdot\aa = \widetilde{\Omb}\times\aa ,
\end{align}
where \( \times \) represents the standard cross-product between 3D vectors, and where \( \widetilde{\Omb} \) is a vector defined through the final
equality. While the use of vectors and cross products is perhaps more common that anti-symmetric matrices, the latter approach simplifies
various arguments and so is preferred within this work.

It is also helpful to define the rotation group's \textit{adjoint representation}, \(\BigAd{}\),   which is a homomorphism from \(\SOThree\)
onto the general linear group of \(\soThree\).  For \( \Omb'\in\soThree \)  and \( \RR\in\SOThree \) we write
\begin{align}
\label{eq:rigid_body:big_ad_defn}
  \BigAd{\RR}\cdot\Omb' \equiv \RR\Omb'\RR^{-1} .
\end{align}
Similarly, the \textit{little adjoint representation}, \( \smallAd{}\), is defined to be the derivative of \( \BigAd{\RR} \) with respect to \( \RR \), and provides a Lie algebra homomorphism of \(\soThree\) on itself. Expanding \( \RR \) about the identity as \( \RR = \id + \Omb \), and with \( \commutator{\cdot}{\cdot} \) the \textit{commutator},
\begin{align}
  \BigAd{\RR}\cdot\Omb' 
  &= \brak{\id + \Omb}\Omb'\brak{\id - \Omb} \nonumber\\
  &= \Omb' + \commutator{\Omb}{\Omb'} + \dots ,
\end{align}
whence we define
\begin{align}
\label{eq:rigid_body:small_ad_defn}
  \smallAd{\Omb}\cdot\Omb' \equiv \commutator{\Omb}{\Omb'} .
\end{align}
The little adjoint operator acts on \( \soLie(3) \) analogously to how the cross product acts on \( \RThree \). Just as \( \mathbf{a} \times \mathbf{a} = 0 \), so \( \smallAd{\Omb}\cdot\Omb = 0 \). Under the homomorphism between \(\soThree\) and \(\RThree\) discussed above, we note
that \( \smallAd{\Omb}\cdot\omb \) is mapped to \( \widetilde{\Omb}\times\widetilde{\omb} \).

Finally, let us ask how torques map from \( \RThree \) into \( \soLie(3) \). Torques commonly take the form
\begin{align}
  \widetilde{\Lext} = \xx\times\mathbf{f} ,
\end{align}
for some position vector \( \xx \) and force \( \mathbf{f} \). We may write the corresponding antisymmetric matrix \( \Lext\in\soLie(3) \) as
\begin{align}
  \Lext = \mathbf{f}\wedge\xx ,
\end{align}
after defining the \textit{skew-symmetric product} \( \wedge : \RThree\times\RThree \rightarrow \soLie(3) \), whose action on two arbitrary vectors \( \mathbf{a} \) and \( \mathbf{b} \) is given by 
\citep[e.g.][]{stone_goldbart}
\begin{align}
\label{eq:rigid_body_wedge_defn}
  \mathbf{a}\wedge\mathbf{b} = 
  \frac{1}{2}\brak{\mathbf{a}\otimes\mathbf{b}-\mathbf{b}\otimes\mathbf{a}} .
\end{align} 
The wedge product of two vectors obeys the relation
\begin{align}
  \BigAd{\RR}\cdot\brak{\aa\wedge\mathbf{b}} = \brak{\RR\cdot\aa}\wedge\brak{\RR\cdot\mathbf{b}} ,
\end{align}
which is analogous to the standard result
\begin{align}
  \RR\cdot\brak{\aa\times\mathbf{b}} = \brak{\RR\cdot\aa}\times\brak{\RR\cdot\mathbf{b}} .
\end{align}
The little adjoint, meanwhile, acts according to
\begin{align}
  \smallAd{\Omb}\cdot\brak{\aa\wedge\mathbf{b}}
  =
  \brak{\Omb\cdot\aa}\wedge\mathbf{b}
  +\mathbf{a}\wedge\brak{\Omb\cdot\mathbf{b}} .
\end{align}

\subsubsection{The translation group}
\label{sec:groups_Tthree}

Finally, we list some properties of the translation group \( \TThree \). The left-action of \( \Trans\in \TThree \) on an arbitrary motion \( \phib \) is
\begin{align}
  \phib \mapsto \Trans\cdot\phib = \phib + \Phib .
\end{align}
It simply translates the motion by some \( \Phib \in \RThree \). Every distinct \( \Trans \) corresponds to translation by some distinct \( \Phib \), so \( \TThree \) is isomorphic to \( \RThree \). Such a constant translation clearly cannot affect velocities, so \( \TThree \) acts on velocities 
as the identity; this follows formally from the fact that
\begin{align}
  \dbyd{t}\braksq{\Trans\cdot\phib(t)}
  =
  \dbyd{t}\braksq{\phib(t) + \Phib}
  =
  \vv(t) .
\end{align}
Next, a one-parameter family \( s \mapsto \Trans(s) \) in the vicinity of the identity acts as
\begin{align}
  \phib \mapsto \Trans(s)\cdot\phib = \phib + s\VV
\end{align}
for arbitrary \( \VV \in \RThree \). Differentiating with respect to \( s \) about \( s = 0 \) shows that an arbitrary element \( \trans \) of the Lie algebra \( \tthree \) acts on \( \phib \) as
\begin{align}
  \phib \mapsto \trans\cdot\phib = \VV .
\end{align}
Thus \( \tthree \) too is isomorphic to \( \RThree \) and has the same action on all motions. We also remark that the adjoint representation of \( \TThree \),
\begin{align}
\label{eq:trans_big_ad_defn}
  \BigAd{\Trans}\cdot\trans
  \equiv
  \Trans\trans\Trans^{-1},
\end{align}
acts on \( \tthree \) as the identity:
\begin{align}
  &
  \brak{\BigAd{\Trans}\cdot\trans}\cdot\phib
  =
  \brak{\Trans\trans\Trans^{-1}}\cdot\phib
  =
  \Trans\cdot\VV 
  =
  \VV
  =
  \trans\cdot\phib 
\end{align}
for an arbitrary motion, so
\begin{align}
  &
  \BigAd{\Trans}\cdot\trans
  =
  \trans .
\end{align}
We have used the fact that \( \trans \)'s action on any motion is just \( \VV \), on which \( \Trans \) acts as the identity.

\subsection{Inner-products}
\label{app_sub:inner_products}

This paper uses several different inner-products. We have the standard inner-product between two Cartesian vectors, which may be defined by
\begin{align}
  \ip{\aa}{\mathbf{b}}
  =
  \frac{1}{4}\brak{
     \norm{\aa+\mathbf{b}}^{2}
    -\norm{\aa-\mathbf{b}}^{2}
  }
\end{align}
for arbitrary \( \aa,\mathbf{b} \in \RThree \). We define the inner product between arbitrary matrices \( \mathbf{A},\mathbf{B} \in \GL(3) \) as
\begin{align}
  \ip{\mathbf{A}}{\mathbf{B}}_{\GL(3)} = \tr{\mathbf{A}\mathbf{B}^{T}} ,
\end{align}
where \( \tr{\cdot} \) denotes the trace operation. This leads us to define the inner-product between elements of \( \sothree \) as
\begin{align}
  \ip{\Omb}{\Omb'}_{\sothree} = - \tr{\Omb\Omb'}
\end{align}
for arbitrary \( \Omb,\Omb' \in \sothree \). Our earlier definition~\eqref{eq:rigid_body_wedge_defn} of the wedge-product means that
\begin{align}
  \ip{\mathbf{b}}{\omb\cdot\aa}
  &\equiv 
  \ip{\omb}{\mathbf{b}\wedge\aa}
  \quad
  \text{for arbitrary } 
  \aa,\mathbf{b}\in \RThree,
  \, 
  \omb\in\soThree ,
\end{align}
which demonstrates a manipulation that we will employ often. While, within this section, we have used subscripts to 
indicate the space over which aninner product is defined, within the main text we leave this implicit
to avoid notational clutter. 


\subsection{Miscellaneous}

We use the notation \( \circ \) to denote composition of two functions. For example,
\begin{align}
    \brak{f' \circ f}(\xx)
    =
    f'\braksq{f(\xx)} .
\end{align}
If functions depend on both space and time, then composition will always be on the spatial variable unless stated otherwise:
\begin{align}
    \brak{f' \circ f}(\xx,t)
    =
    f'\braksq{f(\xx,t),t} .
\end{align}

\section{Linearisation}
\label{sec:app:lin}

\subsection{Linearisation of the tangential slip constraints}

The tangential slip constraint as applied to the motion gives
\begin{align}
  \phib_{-}\z 
  + 
  \epsilon\brak{\FF\z\cdot\chib\f + \uu_{-}\f}
  =
  \phib_{+}\z 
  +
  \epsilon\uu_{+}\f  ,
\end{align}
so at zeroth order
\begin{align}
  \phib_{-}\z = \phib_{+}\z 
\end{align}
while at first order
\begin{align}
\label{eq:chi_defn_first_order}
  \chib\f 
  =
  {\FF\z}^{-1}
  \cdot
  \jump{\uu\f} .
\end{align}
Now, \( \chib\f \in \Tang{\idgen}{\DIFF{\SigFS}} \), so by analogy with the reasoning that led to eq.~\eqref{eq:tangential_slip_constraint_initial}, the linearised tangential slip constraint on the motion is
\begin{align}
\label{eq:tang_slip_constraint_lin}
  \ip{
    {\FF\z}^{-1}
    \cdot 
    \jump{\uu\f}
  }{
    \nhat
  }
  = 0 .
\end{align}

Now consider the tangential slip constraint as applied to the test-functions. We will need the third-rank tensor \( \QFS \) that is defined as
\begin{align}
\label{eq:QFS_defn}
  \QFS\cdot\aa 
  &\equiv
  {\FF\z}^{-T}
  \braksq{
    \nhat
    \otimes
    \aa
  }
  {\FF\z}^{-T}
\end{align}
for arbitrary \( \aa \in \RThree \) (see A18).
It is then a matter of simple algebra to show that
\begin{align}
\label{eq:sdfjhsfkjshfjwfh}
  &
  \intsurf{\SigFS}{
    \pilag
    \ip{
      \ww_{+}\circ\brak{\chib^{-1}}-\ww_{-}
    }{
      J
      \FF^{-T}
      \cdot
      \nhat
    }
  }
  \nonumber\\
  &\qquad
  =
  \intsurf{\SigFS}{
    \pilag
    \ip{
      \jump{\ww}
      -
      \epsilon D\ww_{+}\cdot\chib\f
    }{
      \det\brak{\FF\z + \epsilon D\uu_{-}\f}
      \brak{\FF\z + \epsilon D\uu_{-}\f}^{-T}
      \cdot
      \nhat
    }
  }
  \nonumber\\
  &\qquad
  =
  \intsurf{\SigFS}{
    \pilag
    \ip{
      \jump{\ww}
    }{
      J\z
      {\FF\z}^{-T}
      \cdot
      \nhat
    }
  } 
  -
  \epsilon
  \intsurf{\SigFS}{
    \pilag
    \braksq{
      \ip{D\ww_{+}}{\QFS\cdot\jump{\uu\f}}
      +
      \ip{D\uu_{-}}{\QFS\cdot\jump{\ww}}
    }
  } 
  +
  \order\brak{\epsilon^{2}} ,
\end{align}
where we have used eq.~\eqref{eq:chi_defn_first_order} to produce the terms in \( \QFS \).

Finally, let \( \fullConsSpace{\phib\z} \) be the space of motions \( \uu \) satisfying
\begin{subequations}
\label{eq:lin_func_space_defn}
\begin{align}
  &
  \ip{
    \jump{\uu}
  }{
    {\FF\z}^{-T}\cdot\nhat
  } = 0
  \\
  &
  \intvol{\BB}{\rho\uu} = 0
  \\
  &
  \intvol{\BB}{\rho\uu\wedge\phib\z} .
\end{align}
\end{subequations}
From eqs.~(%
  \ref{eq:euclidean_constraints_linear__u},
  \ref{eq:constraint_new_ang_mom},
  \ref{eq:tang_slip_constraint_lin}%
)
\( \uu\f \) is clearly a member of this space, 
while from eqs.~\eqref{eq:test_function_constraints__exact} and \eqref{eq:sdfjhsfkjshfjwfh} 
so too is \( \ww \) at leading order. From now on we therefore write:
\begin{align}
\label{eq:u_w_lin_same_space}
  \uu\f,\ww \in \fullConsSpace{\phib\z} .
\end{align}

\subsection{Linearisation of the gravitational terms}

We begin with the terms in \( \NNN \): specifically \( \intvol{\BB}{\ip{\NNN}{D\ww}} \) and \( \intsurf{\partial\BB}{\ip{\NNN\cdot\nhat}{\ww}} \). The gravitational stress depends on both \( D\uu\f \) and \( \nabla\zeta\f \), so it linearises to
\begin{align}
  \NNN
  =
  \NNN\z
  +
  \epsilon
  \brak{
    \Gam\cdot D\uu\f 
    +
    \Gamt\cdot \nabla\zeta\f 
  }
  +
  \order\brak{\epsilon^{2}} ,
\end{align}
for some operators \( \Gam \) and \( \Gamt \). Linearising \( \NNN \) in \( \nabla\zeta\f \) is straightforward: from definition~\eqref{eq:grav_stress_defn__explicit} we clearly have
\begin{align}
  \NNN
  =
  \NNN\z 
  +
  \frac{1}{4\pi G} J\z 
  \brak{
     \ip{\gam\z}{\gam\f}\id 
    -\gam\z\otimes\gam\f 
    -\gam\f\otimes\gam\z 
  }
  {\FF\z}^{-T} ,
\end{align}
with \( \gam\f = -{\FF\z}^{-T}\cdot\nabla\zeta\f \). It follows that \( \Gamt \) is given by
\begin{align}
\label{eq:Gamt_defn}
  \Gamt\cdot\aa 
  \equiv
  -\frac{1}{4\pi G} J\z 
  \braksq{
     \ip{{\FF\z}^{-1}\cdot\gam\z}{\aa}\id 
    -\brak{\gam\z\otimes\aa}{\FF\z}^{-1}
    -{\FF\z}^{-T}\brak{\aa\otimes\gam\z} 
  }
  {\FF\z}^{-T} ,
\end{align}
for arbitrary \( \aa \in \RThree \).
Linearising \( \NNN \) in \( D\uu\f \) is no more difficult, but requires more algebra. We write
\begin{align}
  8 \pi G
  \NNN
  =
  8 \pi G
  \NNN\z
  &+
  J\f
  \brak{
     \norm{\gam\z}^{2}\id 
    -2\gam\z\otimes\gam\z
  }
  {\FF\z}^{-T}
  +
  J\z
  \brak{
     \norm{\gam\z}^{2}\id 
    -2\gam\z\otimes\gam\z
  }
  \brak{{\FF}^{-T}}\f
  \nonumber\\
  &+
  2 J\z 
  \brak{
     \ip{\gam\z}{\gam\f}\id 
    -\gam\z\otimes\gam\f 
    -\gam\f\otimes\gam\z 
  }
  {\FF\z}^{-T} 
\end{align}
where here we take \( \gam\f = -\brak{{\FF}^{-T}}\f\cdot\nabla\zeta\z \) and have used the shorthand
\begin{align}
    \brak{{\FF}^{-T}}\f
    =
    -{\FF\z}^{-T}\brak{D\uu\f}^{T}{\FF\z}^{-T} .
\end{align}
It follows that \( \Gam \)'s action on \( D\uu\f \) is
\begin{align}
\label{eq:Gam_defn}
  \Gam\cdot D\uu\f 
  &=
  \NNN\z \braksq{
    \tr{{\FF\z}^{-1}D\uu\f}\id 
    -
    \brak{{\FF\z}^{-1}D\uu\f}^{T}
  }
  \nonumber\\
  &\qquad
  -
  \frac{1}{4\pi G}
  J\z
  \braksq{
     \ip{\gam\z}{\brak{D\uu\f{\FF\z}^{-1}}\cdot\gam\z}\id 
    \mySplitLongBrackets{\qquad\qquad\qquad\qquad\quad}
    -\brak{D\uu\f{\FF\z}^{-1}}^{T}\brak{\gam\z\otimes\gam\z}
    -\brak{\gam\z\otimes\gam\z}\brak{D\uu\f{\FF\z}^{-1}}
  }
  {\FF\z}^{-T} .
\end{align}

\( \Gamt \) appears again in the linearisation of the Poisson term 
\( \brak{4\pi G}^{-1}\intvol{\RThree}{\ip{\aa\cdot\nabla\zeta}{\nabla\potweak}} \). \( \NNN \) and \( \brak{4\pi G}^{-1}\aa\cdot\nabla\zeta \) are the respective first derivatives of \( \brak{8\pi G}^{-1}\ip{\nabla\zeta}{\aa\cdot\nabla\zeta} \) with respect to \( \FF \) and \( \nabla\zeta \), while \( \Gamt \) is the first \textit{mixed} derivative thereof. Having shown that \( \Gamt \) linearly relates \( \NNN \) to \( \nabla\zeta\f \), it follows that \( \Gamt^{T} \) provides the linear relation between the Poisson term and \( D\uu\f \). Thus, we find immediately that
\begin{align}
  \frac{1}{4\pi G}
  \intvol{\RThree}{\ip{\aa\cdot\nabla\zeta}{\nabla\potweak}}
  &=
  \frac{1}{4\pi G}
  \intvol{\RThree}{\ip{\aa\z\cdot\nabla\zeta\z}{\nabla\potweak}}
  +
  \epsilon
  \intvol{\BB}{\ip{
    \Gamt^{T}\cdot D\uu\f  
  }{
    \nabla\potweak
  }}
  \nonumber\\
  &\qquad
  +
  \epsilon
  \frac{1}{4\pi G}
  \intvol{\RThree}{\ip{\aa\z\cdot\nabla\zeta\f}{\nabla\potweak}} ,
\end{align}
where we have used our freedom to redefine the motion and referential potential on \( \RThree/\BB \) to arrange that \( \aa\f \) vanish there, thus converting the second term into an integral over \( \BB \).

\end{document}